\documentclass[a4paper,12pt]{article}
\usepackage{amsfonts,amsthm,amsmath,amssymb,graphicx,youngtab}

\begin{document}

\makeatletter
\@addtoreset{equation}{section}
\makeatother
\renewcommand{\theequation}{\thesection.\arabic{equation}}

\rightline{LYCEN 2013-06, 
WITS-CTP-117}
\vspace{1.8truecm}

\vspace{15pt}


{\large{
\centerline{   \bf Evolution of Yukawa couplings and quark flavour mixings in 2UED models}
\centerline {\bf  }
}}

\vskip.9cm
\thispagestyle{empty} \centerline{
     {\bf Ammar Abdalgabar${}^{a} $\footnote{ {\tt ammar.abdalgabar@students.wits.ac.za}}, A. S. Cornell${}^{a,} $\footnote{ {\tt alan.cornell@wits.ac.za}}, Aldo Deandrea${}^b$\footnote{ {\tt deandrea@ipnl.in2p3.fr}} and Ahmad Tarhini${}^b$\footnote{ {\tt tarhini@ipnl.in2p3.fr}}}}

\vspace{.8cm}
\centerline{{\it ${}^a$ National Institute for Theoretical Physics;}}
\centerline{{\it School of Physics and Centre for Theoretical Physics }}
\centerline{{\it University of the Witwatersrand, Wits, 2050, } }
\centerline{{\it South Africa } }
\vspace{.4cm}
\centerline{{\it ${}^b$ Universit¥e de Lyon, F-69622 Lyon, France}}
\centerline{{\it Universit¥e Lyon 1, CNRS/IN2P3, UMR5822 IPNL}}
\centerline{{\it F-69622 Villeurbanne Cedex, France} }

\vspace{1.4truecm}

\thispagestyle{empty}

\begin{abstract}
The evolution equations of the Yukawa couplings and quark mixings are derived for the one-loop renormalization group equations in 
the two Universal Extra Dimension Models (UED), that is six-dimensional models, compactified in different possible ways to yield 
standard four space-time dimension. Different possibilities for the matter fields are discussed, such as the case of bulk propagating 
or localised brane fields. We discuss in both cases the evolution of the Yukawa couplings, the Jarlskog parameter and the CKM 
matrix elements, and we find that, for both scenarios, as we run up to the unification scale, significant renormalization group 
corrections are present. We also discuss the results of different observables of the five-dimensional UED model in comparison with these
six-dimensional models and the model dependence of the results.
\end{abstract}

\vspace{1cm}


\pagenumbering{arabic}


\setcounter{footnote}{0}

\linespread{1.1}
\parskip 4pt

\section{Introduction}

\par The Standard Model (SM) is an extremely successful theory describing interactions among elementary particles, and its predictions have been experimentally tested to a high level of accuracy. However, there are many parameters which remain unexplained, such as the origin of fermion masses and their associated mixing angles, providing a window to new physics beyond the SM (see for example \cite{Fritzsch:1999ee,GonzalezGarcia:2002dz,Altarelli:2004za} and references therein). Apart from the discovery of the Higgs Boson at the Large Hadron Collider (LHC), another important goal of the LHC is to explore the new physics that may be present at the TeV scale. Among these models those with extra spatial dimensions offer many possibilities for model building and TeV scale physics scenarios which can be constrained or explored. In this context, there have been many attempts to understand the origin of fermion masses and their mixings by making use of Renormalization Group Equations (RGEs) especially for Universal Extra Dimension (UED) models and their possible extensions (see for example \cite{Bhattacharyya:2006ym,Deandrea:2006mh,Cornell:2010sz,Cornell:2011fw,Cornell:2012uw,Cornell:2012qf,Ohlsson:2012hi} and references therein).

\par UED models at the TeV scale \cite{Antoniadis:1990ew} are explored in various configurations, the simplest being the case of one flat extra dimension compactified on $S^1/Z_2$, which is widely studied and constrained since more than a decade \cite{Appelquist:2000nn,Choudhury:2011jk}. In this case each standard model field is accompanied by one massive tower of states, the Kaluza-Klein (KK) particles. If at the boundaries one assumes the same localised interactions, the lightest KK state is stable thanks to a parity, the KK parity, giving a natural  dark matter particle candidate if the particle is neutral \cite{Servant:2002aq,Cheng:2002iz}. An extension of this scenario consist in considering this type of model in two extra dimensions, this extension is non-trivial and brings further insight and is theoretically motivated by specific requirements. As such there are many reasons to study these models, primarily as they provide a dark matter candidate, and suppress the proton decay rate as well as anomaly cancellations from the number of fermion generations being a multiple of three \cite{Dobrescu:2001ae}.

\par Different models with two extra-dimensions have been proposed such as $T^2/Z_2$ \cite{Appelquist:2000nn}, the chiral square $T^2/Z_4$ \cite{Dobrescu:2004zi, Burdman:2005sr}, $T^2/(Z_2 \times Z'_2)$ \cite{Mohapatra:2002ug}, $S^2/Z_2$ \cite{Maru:2009wu}, the flat real projective plane $RP^2$ \cite{Cacciapaglia:2009pa}, the real projective plane starting from the sphere \cite{Dohi:2010vc}. For example in \cite{Cacciapaglia:2009pa} the parity assuring the stability of the dark matter candidate is due to a remnant of the 6-dimensional Lorentz symmetry after compactification as the model has no fixed points (see \cite{Arbey:2012ke} for a detailed discussion). 

\par These compactifications will lead to two towers of new particle states for each standard model particle in the effective 4D theory. In flat geometry the spectrum will be those of a two-dimensional potential well, while for the sphere the spectrum will be the one of angular momentum. As such, in the 4D effective theory there appear infinite towers of massive KK states, with each excitation being specified by two integers, $(j, k)$, called the KK-numbers. For simplicity in this paper we assume that the two extra space-like dimensions have the same size, that is $R_5 = R_6 =R$. This hypothesis is not realistic in all cases as some models are excluded when combining LHC limits and relic dark matter density constraints \cite{Cacciapaglia:2011hx,Cacciapaglia:2012dy,Arbey:2012ke,Kakuda:2013kba}. However this simpler case gives the opportunity to compute in detail the renormalisation group evolution equations (RGE) and study the behaviour of Yukawa couplings and quark mixings in such a scenario. The study of the RGE provides a way by which partial explorations of the physics implications at a high energy scale is possible. By using the RGE, we will study the evolution of the Yukawa couplings and flavour mixings in the quark sector in the charged current, as described by the Cabibbo-Kobayashi-Maskawa (CKM) matrix which has four observable parameters including three mixing angles and one phase. 

\par The form of the CKM matrix in the standard parameterisation is
\begin{equation}
V_{CKM} = \left( {\begin{array}{ccc}
{{V_{ud}}}&{{V_{us}}}&{{V_{ub}}}\\
{{V_{cd}}}&{{V_{cs}}}&{{V_{cb}}}\\
{{V_{td}}}&{{V_{ts}}}&{{V_{tb}}}
\end{array}} \right) = \left( {\begin{array}{ccc}
{{c_{12}}{c_{13}}}&{{s_{12}}{c_{13}}}&{{s_{13}}{e^{ - i{\delta}}}}\\
{ - {s_{12}}{c_{23}} - {c_{12}}{s_{23}}{s_{13}}{e^{i{\delta}}}}&{{c_{12}}{c_{23}} - {s_{12}}{s_{23}}{s_{13}}{e^{i{\delta}}}}&{{s_{23}}{c_{13}}}\\
{{s_{12}}{s_{23}} - {c_{12}}{c_{23}}{s_{13}}{e^{i{\delta}}}}&{ - {c_{12}}{s_{23}} - {s_{12}}{c_{23}}{s_{13}}{e^{i{\delta}}}}&{{c_{23}}{c_{13}}}
\end{array}} \right)\; , \label{eqn:CKM}
\end{equation}
where $s_{12} = \sin\theta_{12}$, $c_{12} = \cos\theta_{12}$ etc. are the sines and cosines of the three mixing angles $\theta_{12}$, $\theta_{23}$ and $\theta_{13}$, and $\delta$ is the CP violating phase. This parameterisation will be used in the rest of the paper.

\par In section \ref{6DUED-description} we introduce the 2UED SM, followed in section \ref{gaugE_EVolution} by a presentation of the evolution equations for the gauge couplings and the comparison between 5D and 6D UED models. In section \ref{beta_Yukawa-ckm} we derive the RGE for the Yukawa couplings and the CKM matrix elements in the 2UED model for both cases, that is for matter fields propagating in the bulk and when they are restricted to the brane. We shall discuss our results and the evolution properties for the physical observables in section \ref{resulTS}, followed by our conclusions in section \ref{conclusiONS}.


\section{The 2UED SM}
\label{6DUED-description}

\par We study a generic model with two universal extra dimensions, where all the SM fields (or some subset) propagate universally in 6D space-time. The space-time coordinate $x_{\mu}(\mu=1,2,3,4)$ denotes the usual Minkowski space, and the two extra spatial dimension coordinates $x_5$ and $x_6$ are compactified. For simplicity we will refer here to the flat extra dimensional notation, however for the purpose of computing renormalisation evolution equation, we will later consider also the case of curved orbifold (the sphere $S^2$ and related orbifolds). 

\subsection{Fermions}

\par The spinor dimension of a fermion $\Psi$ in 6 dimensions is minimally 8 (contrary to 4 minimal components in 4 and 5 dimensions): the Clifford algebra contains six 8$\times$8 gamma matrices $\Gamma^1 \dots \Gamma^6$. Moreover, one can define
\begin{equation}
\Gamma^7 = \Gamma^1 \Gamma^2 \Gamma^3 \Gamma^4 \Gamma^5 \Gamma^6
\end{equation}
and it is possible to define two 6D chiralities
\begin{equation}
P_\pm = \frac{1}{2} \left( 1 \pm \Gamma^7 \right)\,.
\end{equation}
The minimal spinor representation of the Lorentz group are 4-component chiral fermions $\Psi_\pm = P_\pm \Psi$. Each  of the 6D-chiral fields contains two four dimensional Weyl fermions of opposite 4D-chirality. Such considerations are quite general and apply to different models (see for example \cite{Burdman:2005sr,Cacciapaglia:2009pa} for a more detailed discussion of the formalism).

\par For example in case of two flat extra dimensions, the Lagrangian for fermions read :
\begin{eqnarray}
{\cal L}_{Fermions} &=& \int {dx_5} \int {dx_6}\; \left\{ i\bar \psi_{\pm}{\Gamma ^M}{\partial_M} \psi_{\pm} \right\} \\
&=& \int {dx_5} \int {dx_6} \; i \left\{ \bar \psi_{\pm L}{\Gamma^\mu}{\partial_\mu} \psi_{\pm L} +  \bar \psi_{\pm R}{\Gamma^\mu}
{\partial_\mu} \psi_{\pm R} +  \bar \psi_{\pm L}{\Gamma^\pm}{\partial_\mp} \psi_{\pm R} + \bar \psi_{\pm R}{\Gamma^\mp}
{\partial_\pm} \psi_{\pm L} \right\} \;,\nonumber
\end{eqnarray}
where $\Gamma^\pm=\frac{1}{2}(\Gamma^{5}\pm i \Gamma^{6})$ and $\partial\pm=\partial_{5}\pm i \partial_{6}$. 
The way in which 4D chiral zero modes describing the SM fermion are obtained differs in different models.  Most often a quotient of the original symmetry group by a discrete $Z_2$ symmetry is necessary to eliminate one 4D degree of freedom  and allow to have a 4D chiral fermion \cite{Dobrescu:2004zi}, but it can be also obtained directly from the properties of the orbifold as in \cite{Cacciapaglia:2009pa}. Higher massive modes are vector-like fermions.

\subsection{Scalars}

\par The Lagrangian for a scalar field $\Phi$ is
\begin{eqnarray}
{\cal L}_{Scalars} &=& \int {dx_5} \int {dx_6}\;  \Big\{ \partial_\alpha \Phi^\dagger \partial^\alpha \Phi - M^2 \Phi^\dagger \Phi \Big\}\,,
\end{eqnarray}
where $\alpha = 1,\dots 6$ and the corresponding equations of motion are
\begin{equation}
\left( \partial_5^2 + \partial_6^2 + p^2 - M^2 \right) \Phi = 0\,,
\end{equation}
where $p^2 = - \partial_\mu \partial^\mu$. After Fourier decomposition along the extra coordinates, the fields can be written as a sum of KK modes. The wave functions satisfy the above equation with $p^2$ replaced by the mass square of the mode. The solutions of this equation are usual combinations of sines and cosines (with frequencies determined by the periodicity) in flat extra dimensions, while in the case of the 2-sphere inspired orbifolds the solutions are the spherical harmonics. The masses are given by the formula
\begin{equation}
m_{k,l}^2 = M^2 + k^2 + l^2\,,
\end{equation}
and the mass eigenstates can be labelled by their parity assignment with respect to the generators of the symmetry group of the orbifold and by the KK numbers $(k,l)$. 

\subsection{Gauge bosons}

\par The Lagrangian for an Abelian gauge field (also for non-Abelian gauge symmetries at quadratic level) is
\begin{eqnarray}
{\cal L}_{Gauge+GF}&=& \int {dx_5} \int {dx_6}\; \left[ -\frac{1}{4} F^{\alpha\beta} F_{\alpha\beta} -\frac{1}{2\xi} [\partial_\mu A^\mu - \xi (\partial_5 A_5 + \partial_6 A_6)]^2 \right]
\end{eqnarray}
where $\xi$ is the gauge fixing parameter,  and $F_{\alpha \beta} = \partial_\alpha A_\beta - \partial_\beta A_\alpha$. The gauge fixing term eliminates the mixing between $A_\mu$ and the extra polarisation $A_5$ and $A_6$. Once the parities are assigned, the spectrum and wave functions will be the same as for the scalar field (without mass term):
\begin{equation}
m_{k,l}^2 = k^2 + l^2\,.
\end{equation}
In the Feynman-'t~Hooft gauge $\xi = 1$, the equations of motion for $A_5$ and $A_6$ decouple from the rest:
\begin{equation}
(\partial_5^2 + \partial_6^2 - \partial_\mu^2) A_{5,6} = 0
\end{equation}
and the two extra-components of the gauge field can be treated as two independent scalar fields. Spectra and wave functions are again similar to the scalar case, with some additional constraints. Therefore each of the gauge fields has six components \cite{Burdman:2005sr} and decomposes into towers of 4D spin-1 fields and two towers of real scalars belonging to the adjoint representation. The phenomenology of these spinless adjoints has been investigated in detail in \cite{Dobrescu:2007xf}.

\subsection{Yukawa interactions}

\par Yukawa interactions are built in the usual way in terms of 6D fields:
\begin{eqnarray}
{\cal L}_{Yukawa} &=& \int {dx_5} \int {dx_6}\; Y^{ab}\left\{ i\bar \psi_{\pm R}^a \phi \psi_{\mp L}^b \right\}
\;.
\end{eqnarray}
 Any 6D field (fermion/gauge or scalar) $\Phi(x^{\mu}, x^5, x^6)$ can be decomposed as:
\begin{equation}
\Phi(x^{\mu}, x^5, x^6)= \frac{1}{L} \sum_{j,k}{f^{(j,k)}(x^5, x^6) \phi^{(j,k)}(x^{\mu})} \;,
\end{equation}
where in the flat case
\begin{equation}
f^{(j,k)}(x^5, x^6)= \frac{1}{1+ \delta_{0,j}\delta_{0,k}}\left[e^{-\frac{in \pi}{2}} \cos\left(\frac{j x^5 +k x^6}{R}+\frac{n \pi}{2}\right)+
\cos\left(\frac{k x^5 -j x^6}{R}+\frac{n \pi}{2}\right)\right] \;.
\end{equation}
Note that the 4D fields $\phi^{(j,k)}(x^{\mu})$ are the $(j, k)^{th}$ KK modes of the 6D fields $\Phi(x^M)$ and $n$ is an integer whose value is restricted to $0, 1, 2$ or $3$ by the boundary conditions. The zero mode $(j,k =0)$ is allowed only for $n=0$ in the 4D effective theory. These Yukawa interactions will give rise to the usual SM Yukawa interactions plus those related to the towers of KK states.

\subsection{Model dependence of the spectra}

\par In general a fixed value of the KK numbers $(k,l)$ will correspond to a tier of states (including  scalars, but also fermions and gauge bosons) and for each type of particles there will be more than one state (corresponding to the  different possible parities of the orbifold). However not all the possible states will be present as some states may be not possible due to symmetry constraints and boundary conditions. Indeed looking to the typical spectra of the 2UED models we listed in the introduction, one can check that this is the case only if at least one of the two KK numbers $(k,l)$ is equal to zero, while the higher tiers  with $k,l \neq 0$ are fully populated. This is an important observation for the calculation of the RGE, as the  fact that only few of the first KK modes are absent (and which ones depend on the model) has little effect on the numerical results, thus reducing considerably the model dependence of the evolution equations, as we shall see in more detail in the following (see Appendix \ref{Ssq-factor}).

\section{Gauge couplings evolution}
\label{gaugE_EVolution}

\par Armed now with our 2UED model we derive the gauge coupling RGE, where our results agree with \cite{Kakuda:2013kba,Nishiwaki:2011gm} for all matter fields propagating in the bulk. Apart from the SM field contributions, there will be new contributions from the spinless adjoints $A^{(j,k)}_H$, where the calculation is similar to that of the 5D UED model but with an additional  factor of 2 due to 6D gauge field having two extra dimensional components. Note that for the case of all matter fields being restricted to the brane there will be no contributions from the KK excited states of the fermions. The generic structure of the one-loop RGE for the gauge couplings is then given by:
\begin{equation}
16 \pi^2 \frac{d g_i}{d t}= b^{SM}_i g^3_i+\pi \left( S(t)^2-1 \right) b^{6D}_i g^3_i \;,
\label{gauge2UED}
\end{equation}
where $t = \ln (\frac{\mu}{M_Z})$, $S(t) = {e^t}{M_Z}R$, or $S(\mu)=\mu R=\frac{\mu}{M_{KK}}$ for $M_Z < \mu < \Lambda$ ($\Lambda$ is the cut-off scale as shall be discussed in more detail in section \ref{resulTS}). More details about the calculation of the $S^2(t)$ factor can be found in Appendix \ref{Ssq-factor}. The numerical coefficients appearing in equation (\ref{gauge2UED}) are given by:
\begin{equation}
b^{SM}_i=  \left[ \frac{41}{10}, -\frac{19}{6}, -7\right]\;,
\end{equation}
and 
\begin{equation}
b^{6D}_i= \left[\frac{1}{10}, -\frac{13}{2}, -10\right]+\left[\frac{8}{3}, \frac{8}{3}, \frac{8}{3}\right]\eta\;,
\end{equation}
\noindent $\eta$ being the number of generations of fermions propagating in the bulk. Therefore, in the two cases we shall consider, that of all fields propagating in the bulk ($\eta =3$) we have \cite{Cornell:2011fw}:
\begin{equation}
b^{6D}_i= \left[\frac{81}{10}, \frac{3}{2}, -2\right]\;.
\end{equation}
\noindent Similarly, for all matter fields localised to the brane ($\eta =0$) we have:
\begin{equation}
b^{6D}_i= \left[\frac{1}{10}, -\frac{13}{2}, -10\right]\;.
\end{equation}

\begin{figure}[h]
\begin{center}
\includegraphics[width=7cm,angle=0]{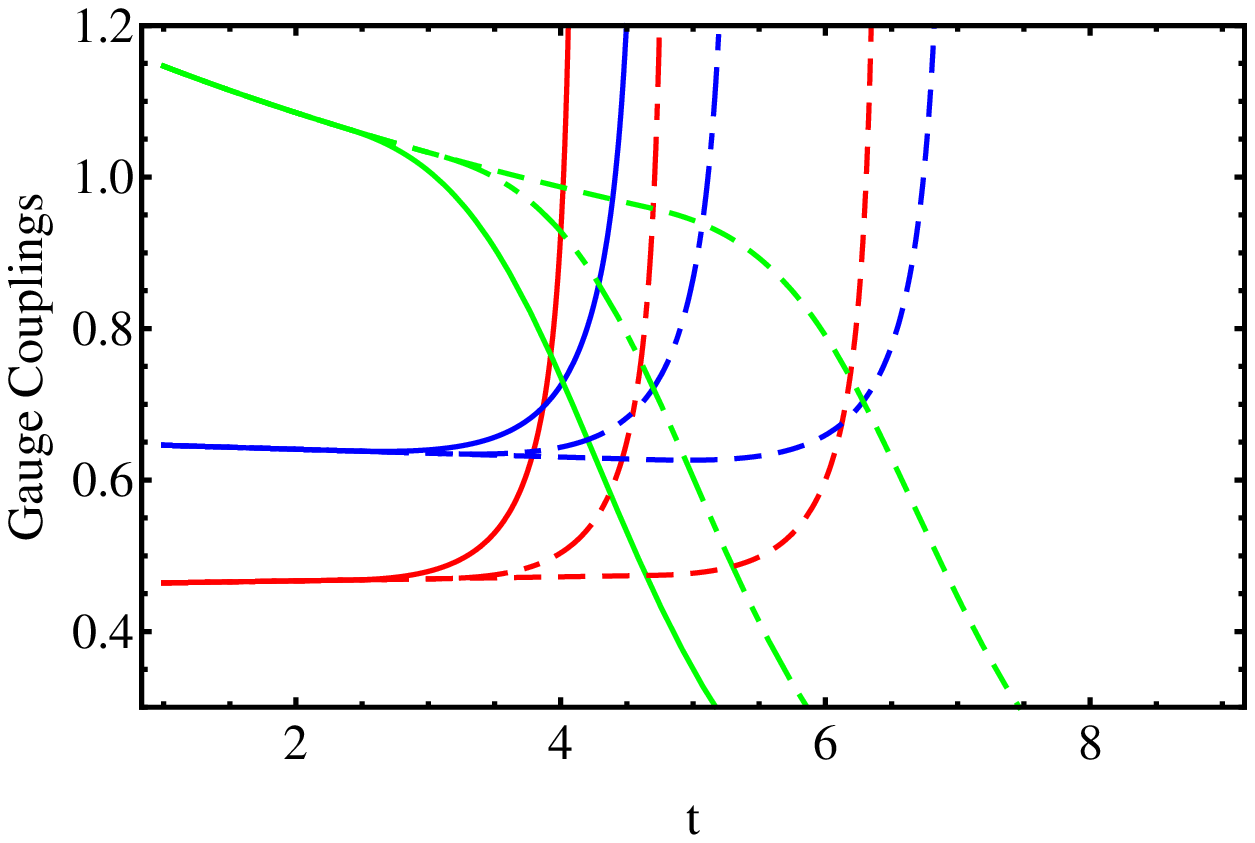} \qquad
\includegraphics[width=7cm,angle=0]{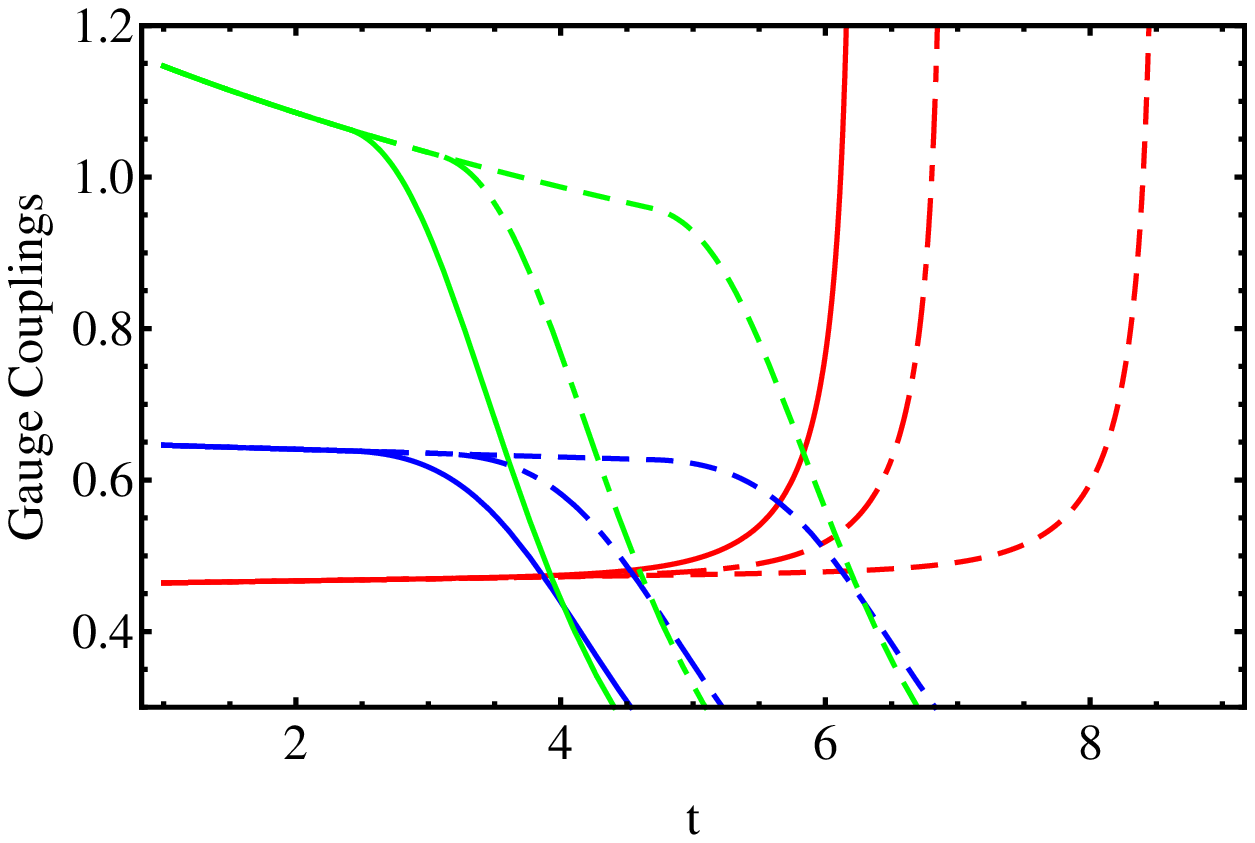}
\end{center}
\caption{ \it (Colour online) The evolution of gauge couplings {$g_1$} (red), {$g_2$} (blue) and {$g_3$} (green), with: in the left panel, all matter fields in the bulk; and the right panel for all matter fields on the brane; for three different values of the compactification scales 1 TeV (solid line), 2 TeV (dot-dashed line) and 10 TeV (dashed line), as a function of the scale parameter {$t$}.}
\label{gauge-2ued}
\end{figure}
\begin{figure}[h]
\begin{center}
\includegraphics[width=7cm,angle=0]{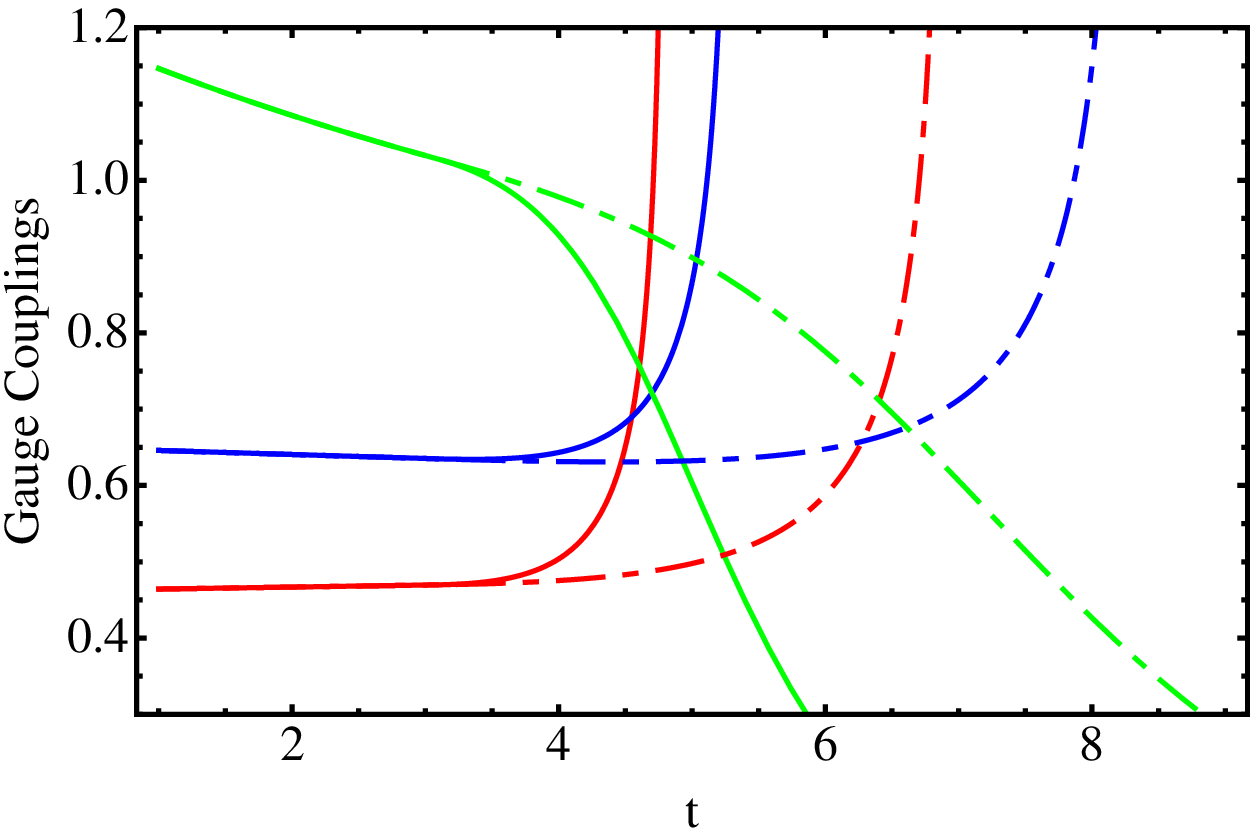} \qquad
\includegraphics[width=7cm,angle=0]{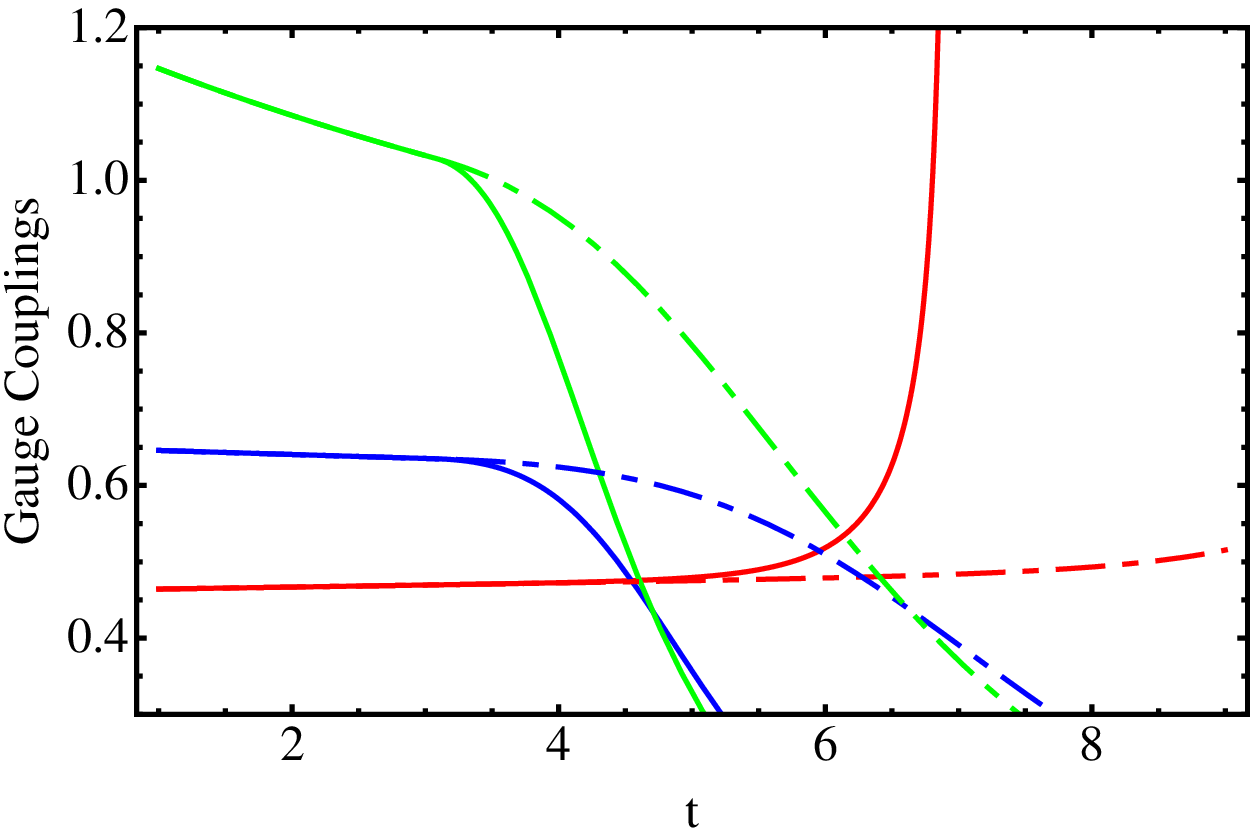}
\end{center}
\caption{ \it (Colour online) Comparison of the gauge coupling evolutions {$g_1$} (red), {$g_2$} (blue), {$g_3$} (green) between the 1UED case (dashed line) and the 2UED case (solid line) with: in the left panel, all matter fields in the bulk; and the right panel for all matter fields on the brane; for a compactification scale of 2 TeV as a function of the scale parameter {$t$}.}
\label{gauge-comparison}
\end{figure}

\par We present in Fig.\ref{gauge-2ued} the evolution of the bulk field and brane localised cases for several choices of compactification scale for the extra-dimension in the 2UED model. We find that there is a difference in the $g_2$ evolution, where it increases in the bulk propagating case and decreases in the brane localised case. We also see that the three gauge coupling constants, as expected in extra-dimensional theories, can unify at some value of $t$ depending on the radius of compactification. As an example, for 1 TeV we see an approximation unification at $t=4$. 
 
\par In Fig.\ref{gauge-comparison} we show for comparison the gauge couplings between the 1UED and 2UED cases for a compactification scale of 2 TeV. From the plots and the discussion in Ref.\cite{Cornell:2012qf}, we see that in both cases the gauge couplings have similar behaviour, however in the 2UED case we have asymptotes at lower $t$ values, that is, a lower energy scale. As such the range of validity for the 2UED is less than the 1UED case, this being due to the $S^2(t)$ factor present in Eq.(\ref{gauge2UED}), there only being a linear dependence on $S(t)$ for the 1UED case.

The solid line (which corresponds to the 2UED case) drops off faster than the dashed line (1UED case) when the gauge couplings decrease with energy scale. For the $g_1$ coupling, it increases faster than in the 2UED case (at $t \sim 6$) with a roughly constant evolution in the 1UED case. As such one can see in the brane case a large difference in the evolution of this coupling, a feature which can distinguish these two models.

\begin{figure}[h]
\begin{center}
\includegraphics[width=7cm,angle=0]{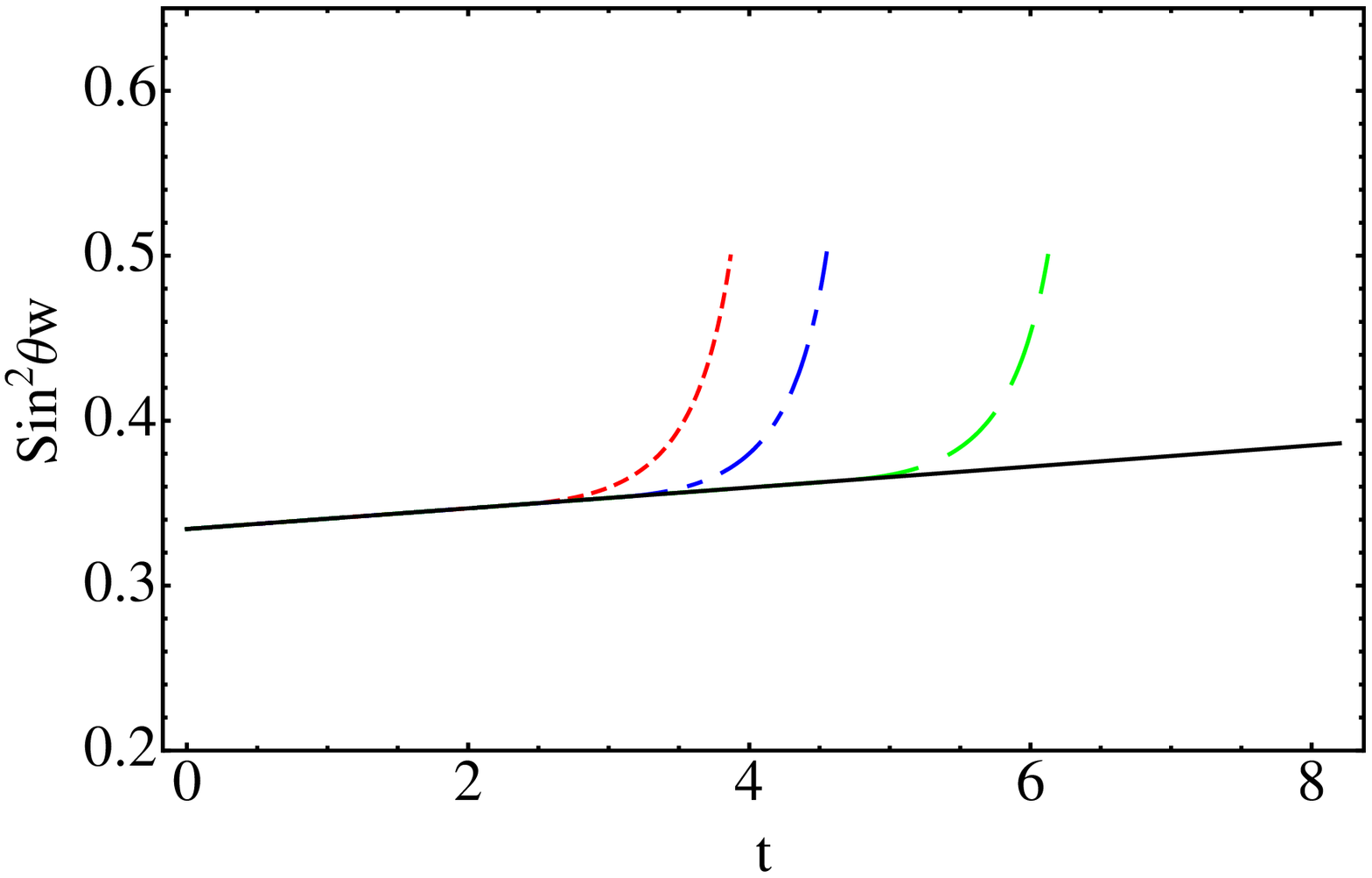} \qquad
\includegraphics[width=7cm,angle=0]{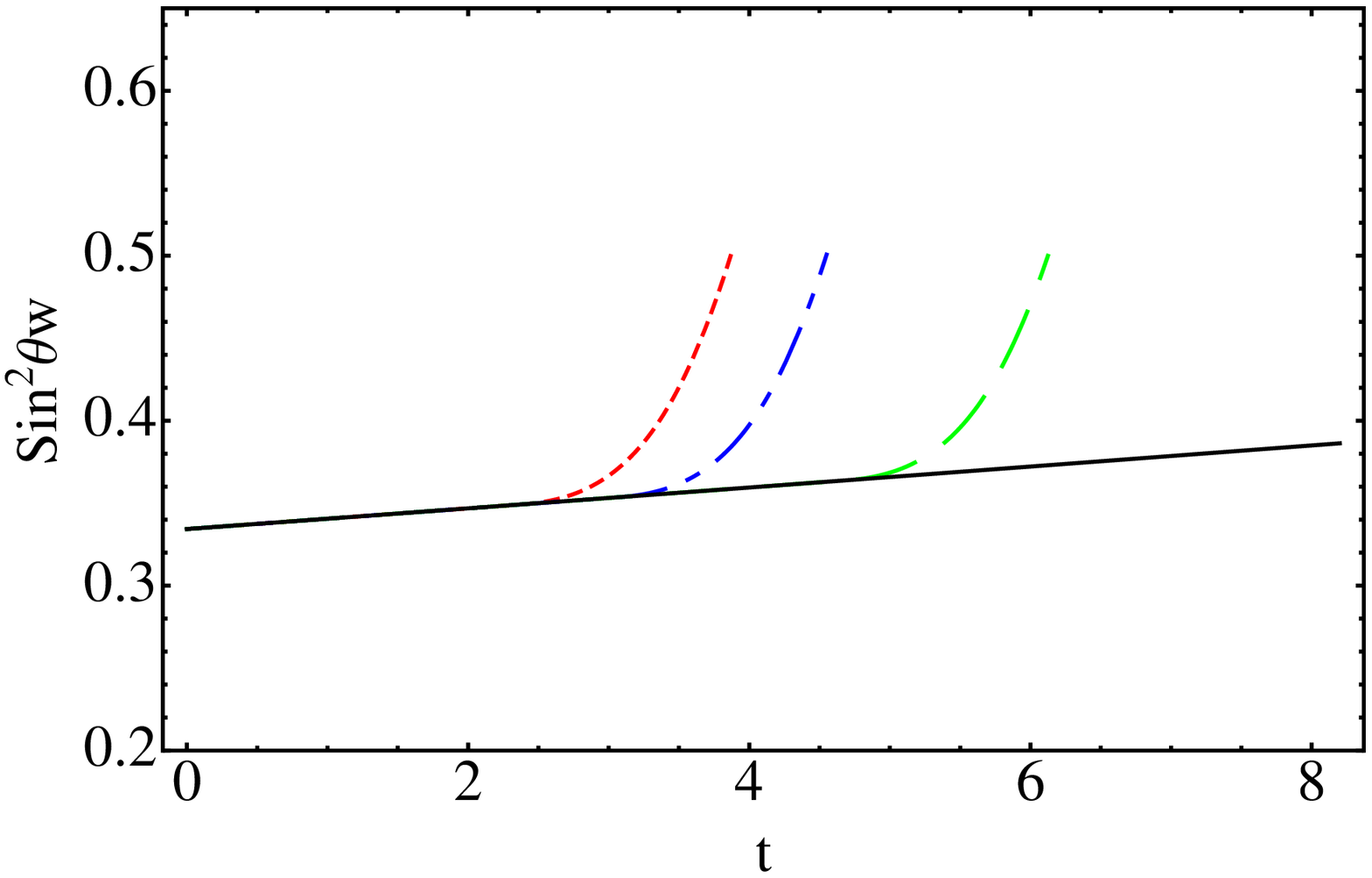}
\end{center}
\caption{ \it (Colour online) The evolution of the Weinberg angle ($\sin^2 \theta_W$) where the solid line represents the SM case with: in the left panel, all matter fields in the bulk; and the right panel for all matter fields on the brane; for three different values of the compactification scales 1 TeV (solid line), 2 TeV (dot-dashed line) and 10 TeV (dashed line), as a function of the scale parameter {$t$}.}
\label{SinTheta-2ued}
\end{figure}

\par In Fig.\ref{SinTheta-2ued} we present the evolution of $\sin^2 \theta_W$ in the 2UED for the bulk and brane cases. Once the KK states begin to contribute the new contributions from the extra-dimensions change the behaviour, that is, it increases until we reach the cut-off scale. One can see that for $R^{-1}=1$ TeV, $\sin^2_{\theta_W}$ can rise to $\sim 0.5$. This result may be useful, at least from a model building perspective, as many extra-dimensional models such as gauge-Higgs unification models in two extra dimensions (see for example \cite{Csaki:2002ur}) predict for many choices of the gauge group large values of $\sin^2 \theta_W$ from a group theory point of view. However, this value is the one expected in the energy range of coupling unification, which once evolved back to the electroweak scale may indeed be close or compatible to the measured value.


\section{Beta functions of the Yukawa couplings and CKM matrix elements in 2UED}
\label{beta_Yukawa-ckm}

\par In the 2UED model the $\beta$-function for the Yukawa couplings can be written as:
\begin{equation}
16 \pi^2 \frac{d Y_{i}}{d t}= \beta_{i}^{SM}+\beta_{i}^{6D}\;, \quad \; \mbox{for}\; i=u, d, e,
\label{general-Yukawa-evolution}
\end{equation}
where $\beta_i^{SM}$ is the SM contribution, and can be found in Refs.\cite{Cornell:2010sz, Cornell:2012qf}. The $\beta_i^{6D}$ are the contributions from the excited KK modes and $S(t)$ is the number of KK levels that fulfils  the inequality $1\leq j^2 + k^2 \leq (\frac{\mu}{M_{KK}})^2$ in this general 2UED model. Recall that $\mu$ is the energy scale and $M_{KK} = R^{-1}$ is the energy for which the first KK mode is generated.

\subsection{Bulk case}

\par For all matter fields propagating in the bulk, we get:
 \begin{eqnarray}
\beta_{u}^{6D} &=&  \pi(S(t)^2-1)Y_u \left[-\frac{32}{3}g_3^2 -\frac{3}{2}g^2_2 -\frac{5}{6} g^2_1+3 (Y^{\dagger}_u Y_u-Y^{\dagger}_d Y_d)\right. \nonumber\\
&&\hspace{3cm}+\left. 2 Tr (3Y^{\dagger}_u Y_u+3Y^{\dagger}_d Y_d+Y^{\dagger}_e Y_e) \right]\; , \\
\beta_{d}^{6D} &=& \pi(S(t)^2-1) Y_d \left[-\frac{32}{3}g_3^2 -\frac{3}{2}g^2_2 -\frac{1}{30} g^2_1+3 (Y^{\dagger}_d Y_d-Y^{\dagger}_u Y_u) \right. \nonumber\\
&&\hspace{3cm}+\left. 2 Tr (3Y^{\dagger}_u Y_u+3Y^{\dagger}_d Y_d+Y^{\dagger}_e Y_e) \right] \; , \\
\beta_{e}^{6D} &=& \pi(S(t)^2-1) Y_e \left[ -\frac{3}{2}g^2_2 -\frac{27}{10} g^2_1+3 Y^{\dagger}_e Y_e \right. \nonumber\\
&&\hspace{3cm}+\left. 2 Tr (3Y^{\dagger}_u Y_u+3Y^{\dagger}_d Y_d+Y^{\dagger}_e Y_e) \right] \; .
\label{Beta-Yukawa-Bulk}
\end{eqnarray}
Note that the coupling constant $g_1$ is chosen to follow the conventional $SU(5)$ normalisation.

\par These Yukawa coupling matrices can be diagonalised by using two unitary matrices {$U$} and {$V$}, where 
\begin{equation}
 UY^{\dagger}_u Y_u U^{\dagger}= \mathrm{diag}(f^2_u,f^2_c,f^2_t)\;, \qquad VY^{\dagger}_d Y_d V^{\dagger}
 = \mathrm{diag}(h^2_d,h^2_s,h^2_b)\; ,
 \end{equation}
in which $f^2_u,f^2_c,f^2_t$ and $h^2_d,h^2_s,h^2_b$ are the eigenvalues of $Y^{\dagger}_u Y_u$ and $Y^{\dagger}_d Y_d$ respectively. As such we obtain the following two relations:
\begin{eqnarray}
 \nonumber 16 \pi^2 \frac{df^2_i}{dt} &=& f^2_i \left[ 2\left( 2\pi (S(t)^2-1 )+1\right)T -2G_u+6 \left( \pi (S(t)^2-1 )+1\right)f_i^2\right. \\ 
 \nonumber &&\hspace{0.6cm}-\left. 6\left( \pi (S(t)^2-1 )+1\right) \sum_{j}{h^2_{j}|V_{i j}|^2} \right] \;, \\
  \nonumber 16 \pi^2 \frac{dh^2_j}{dt} &=& h^2_j \left[ 2\left( 2\pi (S(t)^2-1 )+1\right)T -2G_d+6 \left( \pi (S(t)^2-1) +1\right)h_i^2\right. \\ 
 &&\hspace{0.6cm}-\left. 6\left( \pi (S(t)^2-1 )+1\right) \sum_{i}{f^2_{i}|V_{i j}|^2} \right] \; ,
 \label{yuKAwaSq-evolution}
 \end{eqnarray}
where $i=(u, c, t)$ and $j=(d, s, b)$. Similarly the variation of the lepton Yukawa couplings $y^2_a$ ($a=e, \mu, \tau$) is 
\begin{eqnarray}
 16 \pi^2 \frac{dy^2_a}{dt} = y^2_a\left[ 2\left( 2\pi (S(t)^2-1 )+1\right)T -2G_e +6\left( \pi (S(t)^2-1 )+1\right)y_a^2 \right] \; .
 \label{lePtonSq-evolution}
\end{eqnarray}
In Eqs.(\ref{yuKAwaSq-evolution}, \ref{lePtonSq-evolution}) where have used the following expressions:
\begin{eqnarray}
G_u &=& 8g_3^2 +\frac{9}{4}g^2_2 +\frac{17}{20} g^2_1+\pi\left(S(t)^2-1\right)\left(\frac{32}{3}g_3^2 +\frac{3}{2}g^2_2 +\frac{5}{6} g^2_1\right)\;, \nonumber \\
G_d &=& 8g_3^2 +\frac{9}{4}g^2_2 +\frac{1}{4} g^2_1+\pi\left(S(t)^2-1\right)\left(\frac{32}{3}g_3^2 +\frac{3}{2}g^2_2 +\frac{1}{30} g^2_1\right)\;, \nonumber \\ 
G_e &=& \frac{9}{4}g^2_2 +\frac{9}{4} g^2_1+\pi\left(S(t)^2-1\right)\left(\frac{3}{2}g^2_2 +\frac{27}{10} g^2_1\right)\;,\nonumber\\
T&=& Tr (3Y^{\dagger}_u Y_u+3Y^{\dagger}_d Y_d+Y^{\dagger}_e Y_e)\nonumber\;.
\end{eqnarray}
 
\par The CKM matrix is then obtained upon diagonalisation of the quark mass matrices, {$V_{CKM} = U V^{\dagger}$. The variation of the CKM matrix and its evolution equation for all matter fields in the bulk is:
\begin{eqnarray}
16 \pi^2 \frac{dV_{ik}}{dt}&=& -6 \left( \pi (S(t)^2-1 )+1\right) \left[\sum_{m, j \neq i}{ \frac{f_i^2+f_j^2}{f_i^2-f_j^2} h_{m}^2 V_{im} V^\ast_{j m}}V_{j k}\right. \nonumber \\
&&\hspace{4.1cm}+\left. \sum_{j, m\neq k}{\frac{h_{k}^2+h_{m}^2}{h_{k}^2-h_{m}^2} f_j^2 V^\ast_{jm} V_{j k}V_{i m}}\right] \;.
\label{Vub_evolution_BULK}
\end{eqnarray}

\par The RGEs for the squares of the absolute values of the CKM matrix elements, i.e. the rephasing invariant variables, can now be calculated as:
\begin{eqnarray}
16 \pi^2 \frac{d|V_{ij}|^2}{dt}&=& 2\left( \pi (S(t)^2-1 )+1\right) \left[3|V_{i j}|^2\left(f^2_i+h^2_j-\sum_{k}{f^2_{k}|V_{k j}|^2}-\sum_{k}{h^2_{k}|V_{i k}|^2}\right) \right.\nonumber \\
&&\hspace{2.8cm}-\left. 3 f^2_i\sum_{k\neq i} \frac{1}{f_i^2-f_k^2}\left(2h^2_j|V_{k j}|^2 |V_{i j}|^2+\sum_{l\neq j}h^2_{l}V_{iklj}\right) \right. \nonumber\\ 
&&\hspace{2.8cm}-\left. 3h^2_j\sum_{l\neq j} \frac{1}{h_j^2-h_l^2}\left(2f^2_i|V_{il}|^2 |V_{i j}|^2+\sum_{k\neq i}f^2_{k}V_{iklj}\right)\right]\;, \label{Vijsq-bulk}
\end{eqnarray}
\noindent where 
\begin{equation}
V_{iklj}= 1-|V_{i l}|^2-|V_{k l}|^2-|V_{k j}|^2-|V_{i j}|^2-|V_{i l}|^2 |V_{k j}|^2- |V_{k l}|^2 |V_{i j}|^2 \;.
\end{equation}

\subsection{Brane case}

\par We shall now consider the case of brane localised matter fields for Yukawa couplings in a 6D model. In this case there are no contributions from the KK excited states of the fermions to the Yukawa couplings, in which case we obtain:
 \begin{eqnarray}
\beta_{u}^{6D} &=& 4\pi(S(t)^2-1)Y_u \left[-8g_3^2 -\frac{9}{4}g^2_2 -\frac{17}{20} g^2_1+\frac{3}{2} (Y^{\dagger}_u Y_u-Y^{\dagger}_d Y_d) \right]\; ,\\
\beta_{d}^{6D} &=&  4\pi(S(t)^2-1) Y_d \left[-8g_3^2 -\frac{9}{4}g^2_2 -\frac{1}{4} g^2_1+\frac{3}{2} (Y^{\dagger}_d Y_d-Y^{\dagger}_u Y_u)  \right] \; , \\
\beta_{e}^{6D} &=& 4\pi(S(t)^2-1) Y_e \left[ -\frac{9}{4}g^2_2 -\frac{9}{4} g^2_1+\frac{3}{2} Y^{\dagger}_e Y_e  \right] \;.
\label{Beta-Yukawa-Brane}
\end{eqnarray}

\par By imposing the unitary transformation on both sides of the evolution equations of $Y^{\dagger}_u Y_u$ and $Y^{\dagger}_d Y_d$, we derive the RGEs for the eigenvalues of the square of these Yukawa coupling matrices as follows:
\begin{eqnarray}
  \nonumber 16 \pi^2 \frac{df^2_i}{dt} &=& f^2_i\left[ 2T - 2 G_u +3\left( 4\pi(S(t)^2-1)+1 \right)f_i^2 -3\left( 4\pi(S(t)^2-1)+1 \right) \sum_{j}{h^2_{j}|V_{i j}|^2} \right] \; , \\ \nonumber
 16 \pi^2 \frac{dh^2_j}{dt} &=& h^2_j\left[ 2T - 2 G_d +3\left( 4\pi(S(t)^2-1)+1 \right)h_j^2 -3\left( 4\pi(S(t)^2-1)+1 \right) \sum_{i}{f^2_{i}|V_{i j}|^2} \right] \; , \\ 
16 \pi^2 \frac{dy^2_a}{dt} &=& y^2_a\left[ 2T - 2 G_e +3\left( 4\pi(S(t)^2-1)+1 \right)y_a^2 \right] \; ,
\end{eqnarray}
\noindent where 
\begin{eqnarray}
G_u &=&  8g_3^2 +\frac{9}{4}g^2_2 +\frac{17}{20} g^2_1+ 4\pi\left(S(t)^2-1\right)\left(8g_3^2 +\frac{9}{4}g^2_2 +\frac{17}{20} g^2_1\right)\;, \nonumber \\
G_d &=&  8g_3^2 +\frac{9}{4}g^2_2 +\frac{1}{4} g^2_1 +  4\pi\left(S(t)^2-1\right) \left(8g_3^2 +\frac{9}{4}g^2_2 +\frac{1}{4} g^2_1\right)\;, \nonumber \\ 
G_e &=& \frac{9}{4}g^2_2 +\frac{9}{4} g^2_1 +  4\pi\left(S(t)^2-1\right)\left(\frac{9}{4}g^2_2 +\frac{9}{4} g^2_1\right)\;,\nonumber\\
T&=& Tr (3Y^{\dagger}_u Y_u+3Y^{\dagger}_d Y_d+Y^{\dagger}_e Y_e)\nonumber\;.
\end{eqnarray}
Consequently the CKM running of the quark flavour mixing matrix ($\displaystyle 16 \pi^2 \frac{dV_{ik}}{dt}$ and $\displaystyle 16 \pi^2 \frac{d|V_{i j}|^2}{dt}$) for all matter fields on the brane it is the same as in the bulk case except that the prefactor $2 \left( \pi (S(t)^2-1 )+1\right)$ is replaced by $\left( 4\pi(S(t)^2-1)+1 \right)$ in Eqs.(\ref{Vub_evolution_BULK}, \ref{Vijsq-bulk}).
 
\section{Numerical results and discussions}
\label{resulTS}

\par For our numerical calculations we assume that the fundamental scale is not far from the range of the LHC and set the compatification radii to be {$R^{-1}= 1$} TeV, 2 TeV and 10 TeV. Only some selected plots will be shown and we will comment on the other similar cases not explicitly presented. We quantitatively analyse these quantities in the 2UED model with the initial values adopted at the $M_Z$ scale as: for the gauge couplings $\alpha_1$($M_Z$) = 0.01696, $\alpha_2$($M_Z$) = 0.03377, and $\alpha_3$($M_Z$) = 0.1184; for the fermion masses $m_u$($M_Z$) = 1.27 MeV, $m_c$($M_Z$) = 0.619 GeV, $m_t$($M_Z$) = 171.7 GeV, $m_d$($M_Z$) = 2.90 MeV, $m_s$($M_Z$) = 55 MeV, $m_b$($M_Z$) = 2.89 GeV, $m_e$($M_Z$) = 0.48657 MeV, $m_\mu$($M_Z$) = 102.718 MeV, and $m_\tau (M_Z) = 1746.24$ MeV as in \cite{Cornell:2011fw, Xing:2007fb}.

\begin{figure}[h]
\begin{center}
\includegraphics[width=7cm,angle=0]{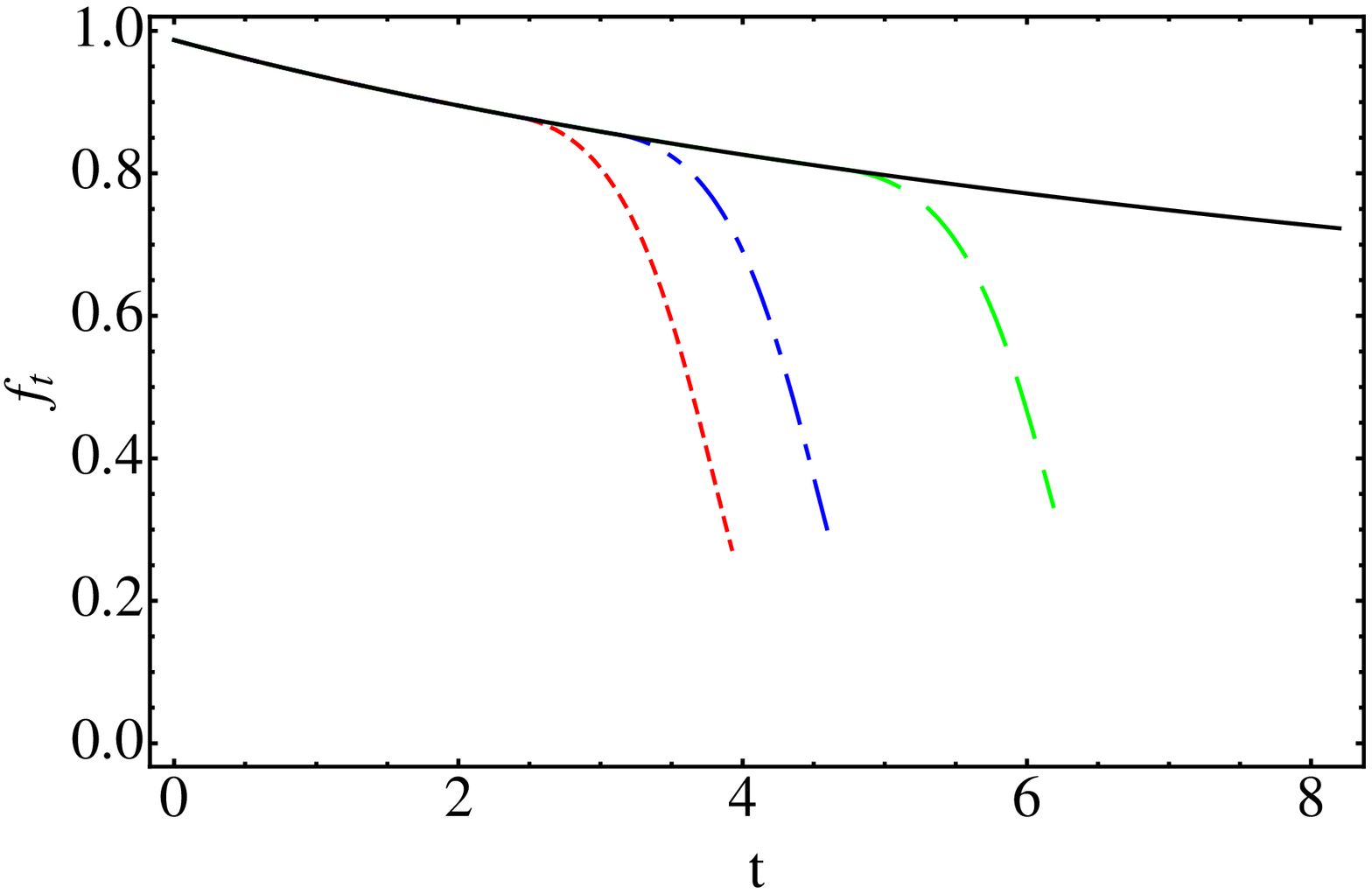} \qquad
\includegraphics[width=7cm,angle=0]{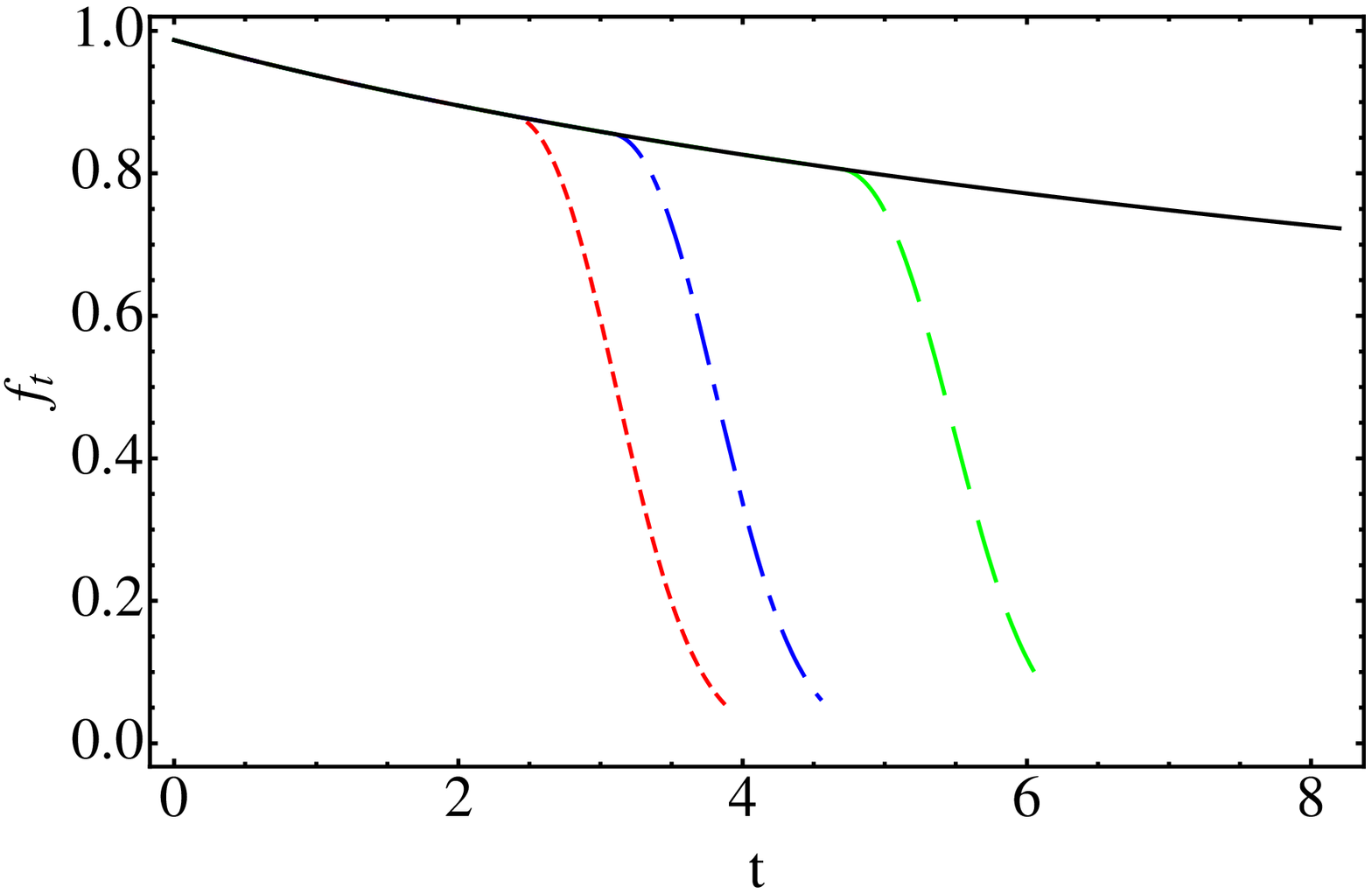}
\end{center}
\caption{ \it (Colour online) The evolution of top Yukawa coupling {$f_t$} where the solid line represents the SM case with: in the left panel all matter fields in the bulk; and the right panel for all matter fields on the brane; for three different values of the compactification scales 1 TeV (solid line), 2 TeV (dot-dashed line) and 10 TeV (dashed line), as a function of the scale parameter {$t$}.}
\label{ft-2ued}
\end{figure}
\begin{figure}[h]
\begin{center}
\includegraphics[width=7cm,angle=0]{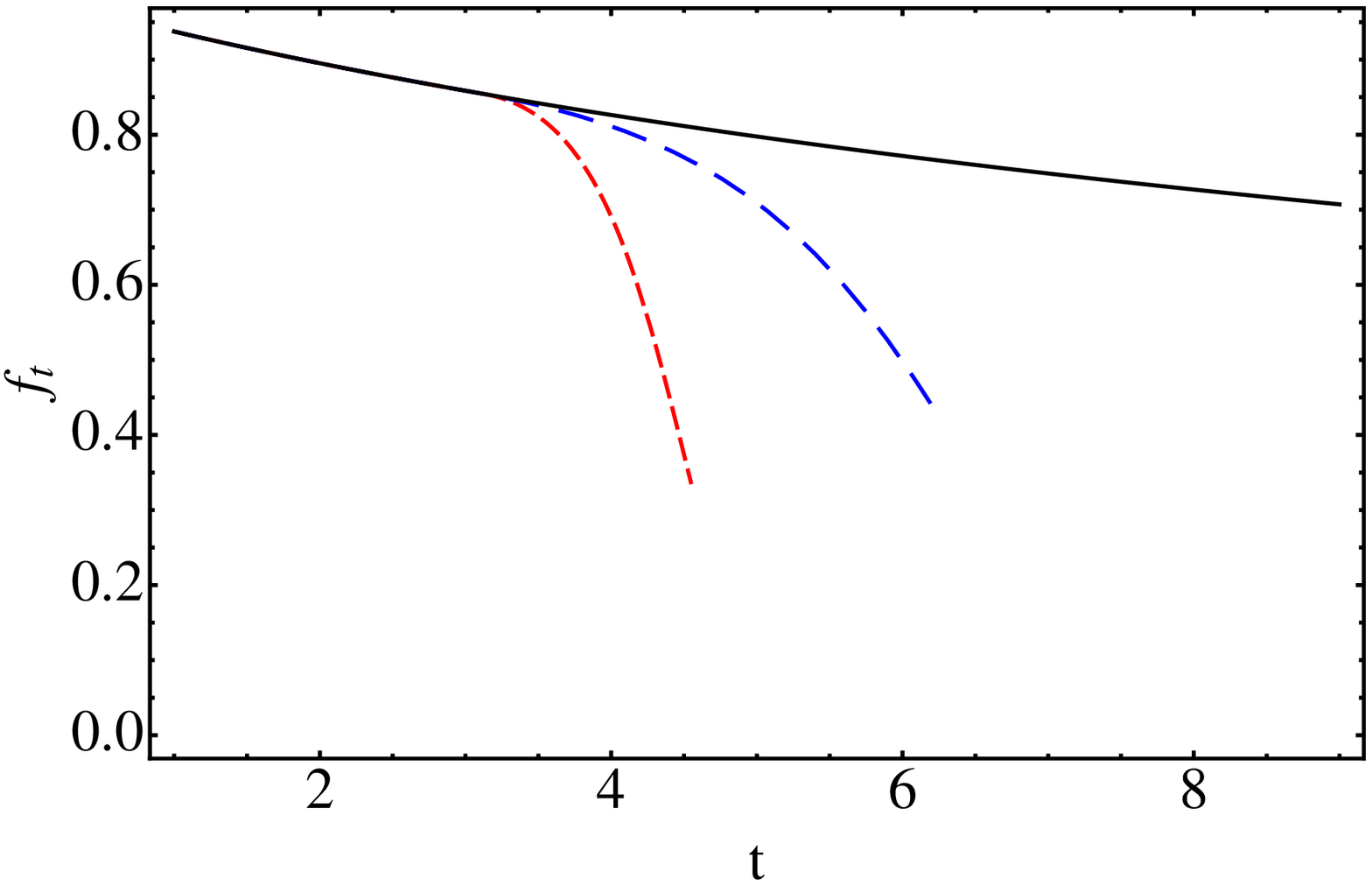} \qquad
\includegraphics[width=7cm,angle=0]{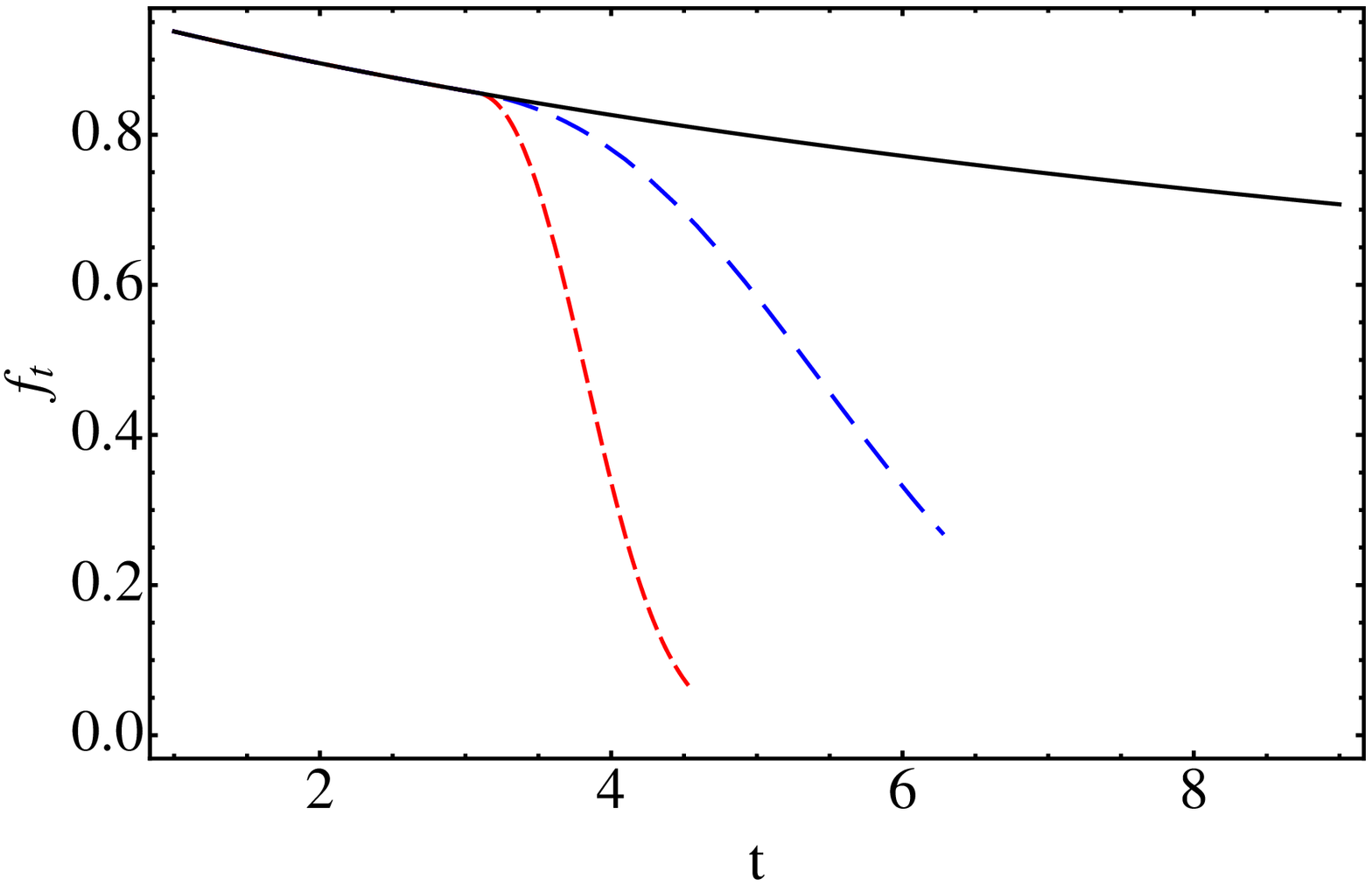}
\end{center}
\caption{ \it (Colour online) Comparison of the top Yukawa coupling evolution between the 1UED case (blue) and the 2UED case (red), where the solid line represent the SM case with: in the left panel all matter fields in the bulk; and the right panel for all matter fields on the brane; for a compactification scale of 2 TeV as a function of the scale parameter {$t$}.}
\label{ft-comparison}
\end{figure}

\par Once the first KK threshold is reached, the contributions from the KK states become more and more significant due to the power law running where the second term on the right hand side of Eq.(\ref{general-Yukawa-evolution}) depends explicitly on the cut-off, which has quantum corrections to the beta functions at each massive KK excitation level. Therefore, the running of the Yukawa couplings deviates from their normal orbits and starts to evolve faster. Similarly, for the Yukawa couplings, where we show in Fig.\ref{ft-2ued} the evolution of the top Yukawa coupling in the 2UED case, the cases of bulk fields and brane localised fields for different radii of compactification, the Yukawa couplings decrease when the first two towers of KK states are reached (that is, when $t > \ln \left(\frac{1}{M_Z R}\right)$). However, as the cut-off is reached quickly in the bulk case, the resulting decrease is of 50\% from the initial value, while in the brane case, the top Yukawa coupling can reach a smaller value by running in a larger energy range (a decrease of about 90\% from the initial value). This is due to the theory being valid to a higher cut-off scale in the brane case.
 
\par A comparison between the 1UED and 2UED cases for the evolution of the top Yukawa coupling is shown in Fig.\ref{ft-comparison} where a rapid decrease appears in the 2UED case, due to the presence of two towers of KK states, which manifest in Eqs.(\ref{Beta-Yukawa-Bulk}, \ref{Beta-Yukawa-Brane}) as the $S^2(t)$ factor. In the 1UED case we have one tower of KK states and a linear dependence of $S(t)$, so we observe that $f_t$ decreases less rapidly than the 2UED case. Note that the evolution of other Yukawa couplings have similar behaviours of decreasing when the quantum corrections from the extra-dimensions set in. 
  
\par We should explicitly state, at this point, that the cut-offs used for the bulk and brane cases in both five and six dimension (1UED and 2UED) are summarised in Tab.\ref{cutoff-tab}. From this we see that the theory in 6D is valid only up to a smaller value of $t$ than the 5D case, where beyond these cut-offs the model would be superseded by new physics. These values correspond to the point where $g_1=g_2$.
 
\begin{table}[!htb]
\caption{ \it The cut-offs in 5D and 6D for both bulk and brane cases for the three compactification radii $R^{-1}=$1, 2 and 10 TeV, where $t=\ln \left(\frac{\mu}{M_Z}\right)$.}
\begin{center}
\begin{tabular}{|c|c|c|c|}
\hline
Scenarios & $t(R_1)$ & $t(R_2)$ & $t(R_3)$\\ \hline
Brane and Bulk 5D (1UED) & 5.61 & 6.27 & 7.81 \\
\hline
Brane and Bulk 6D (2UED) & 3.87 & 4.55 & 6.12 \\
 \hline
\end{tabular}
\label{cutoff-tab}
\end{center}
\end{table}
\begin{figure}[h]
\begin{center}
\includegraphics[width=7cm,angle=0]{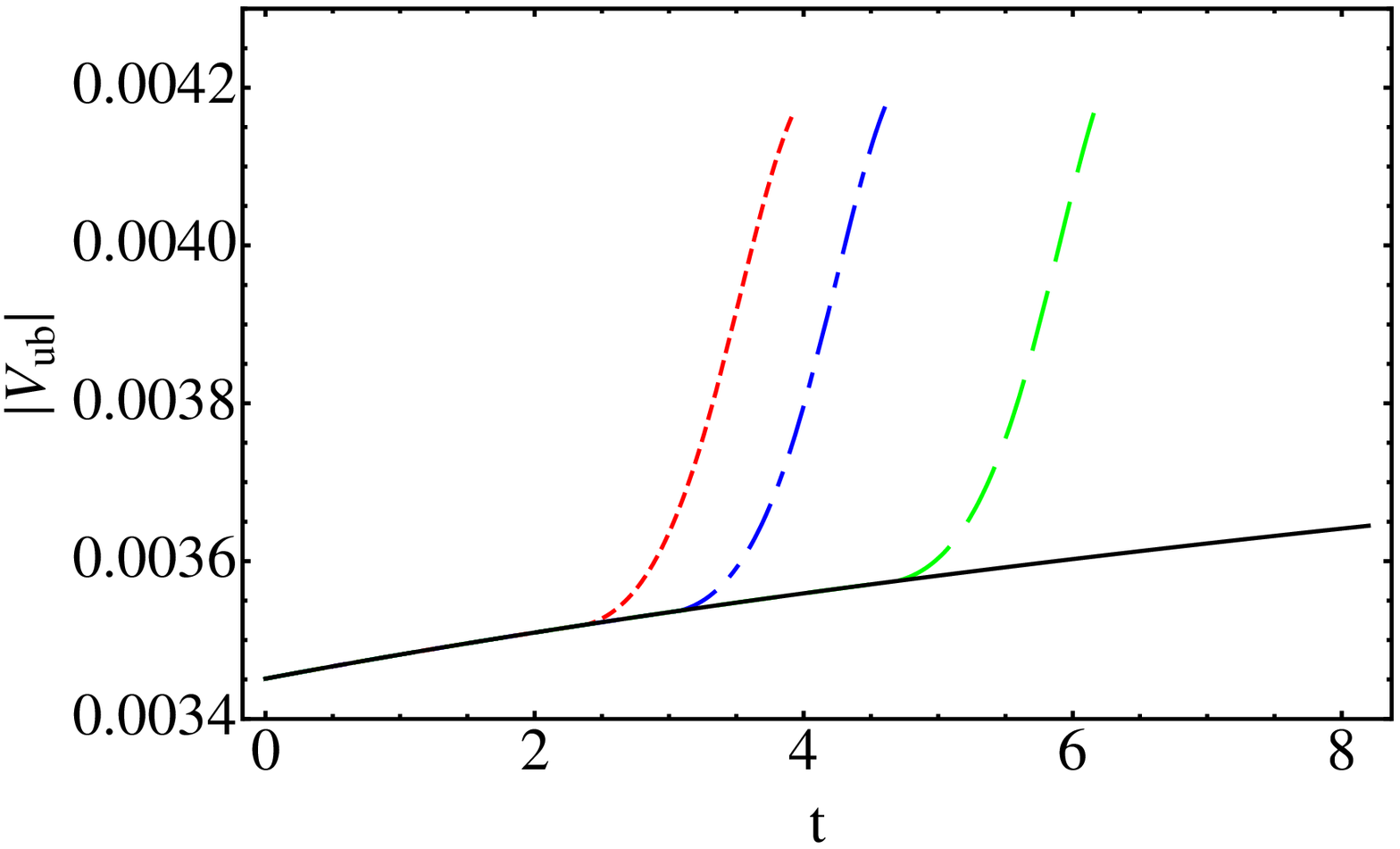} \qquad
\includegraphics[width=7cm,angle=0]{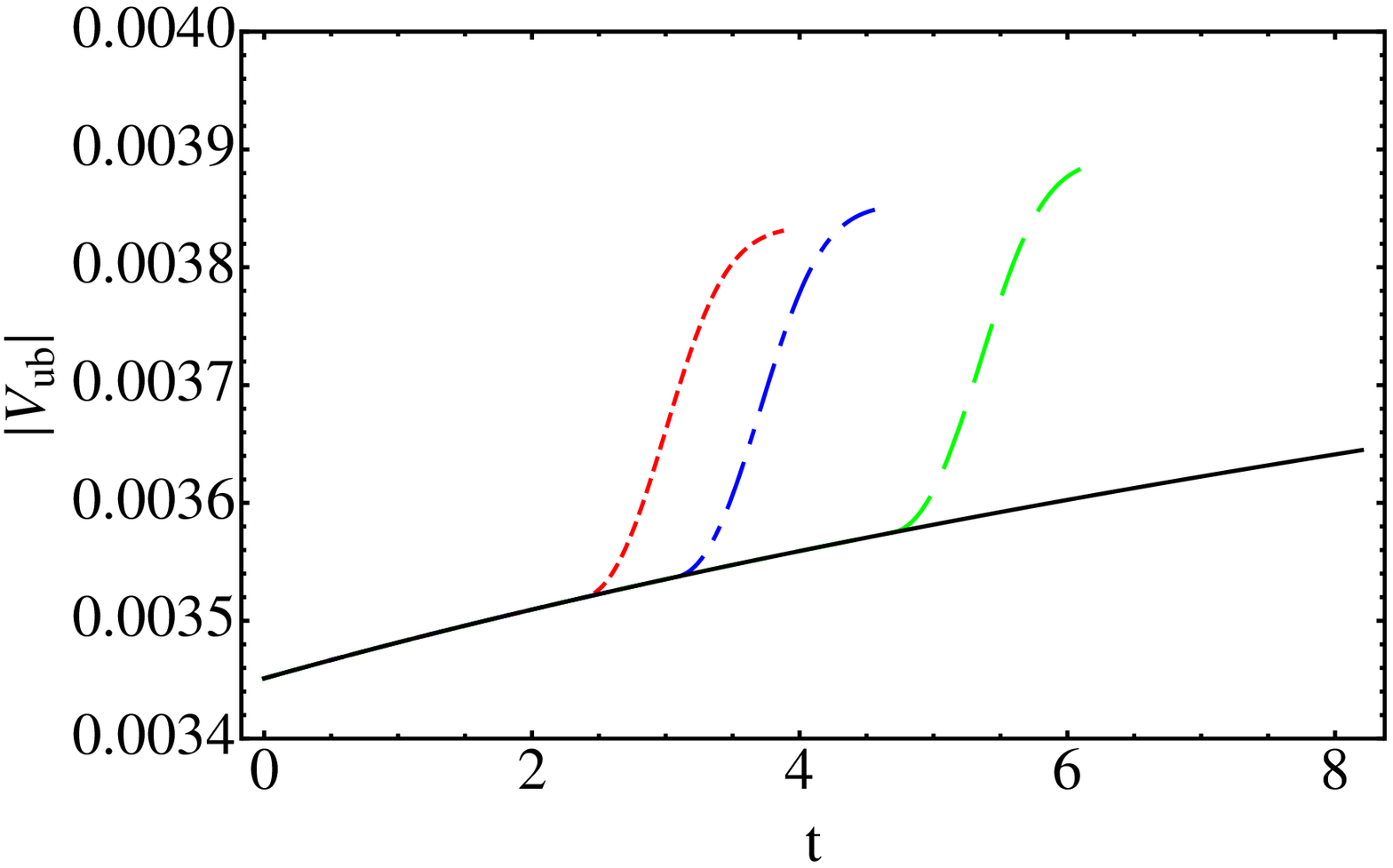}
\end{center}
\caption{ \it (Colour online) The evolution of CKM element {$|V_{ub}|$} where the solid line represents the SM case with: in the left panel all matter fields in the bulk; and the right panel for all matter fields on the brane; for three different values of the compactification scales 1 TeV (solid line), 2 TeV (dot-dashed line) and 10 TeV (dashed line), as a function of the scale parameter {$t$}.}
\label{Vub-2ued}
\end{figure}
\begin{figure}[h]
\begin{center}
\includegraphics[width=7cm,angle=0]{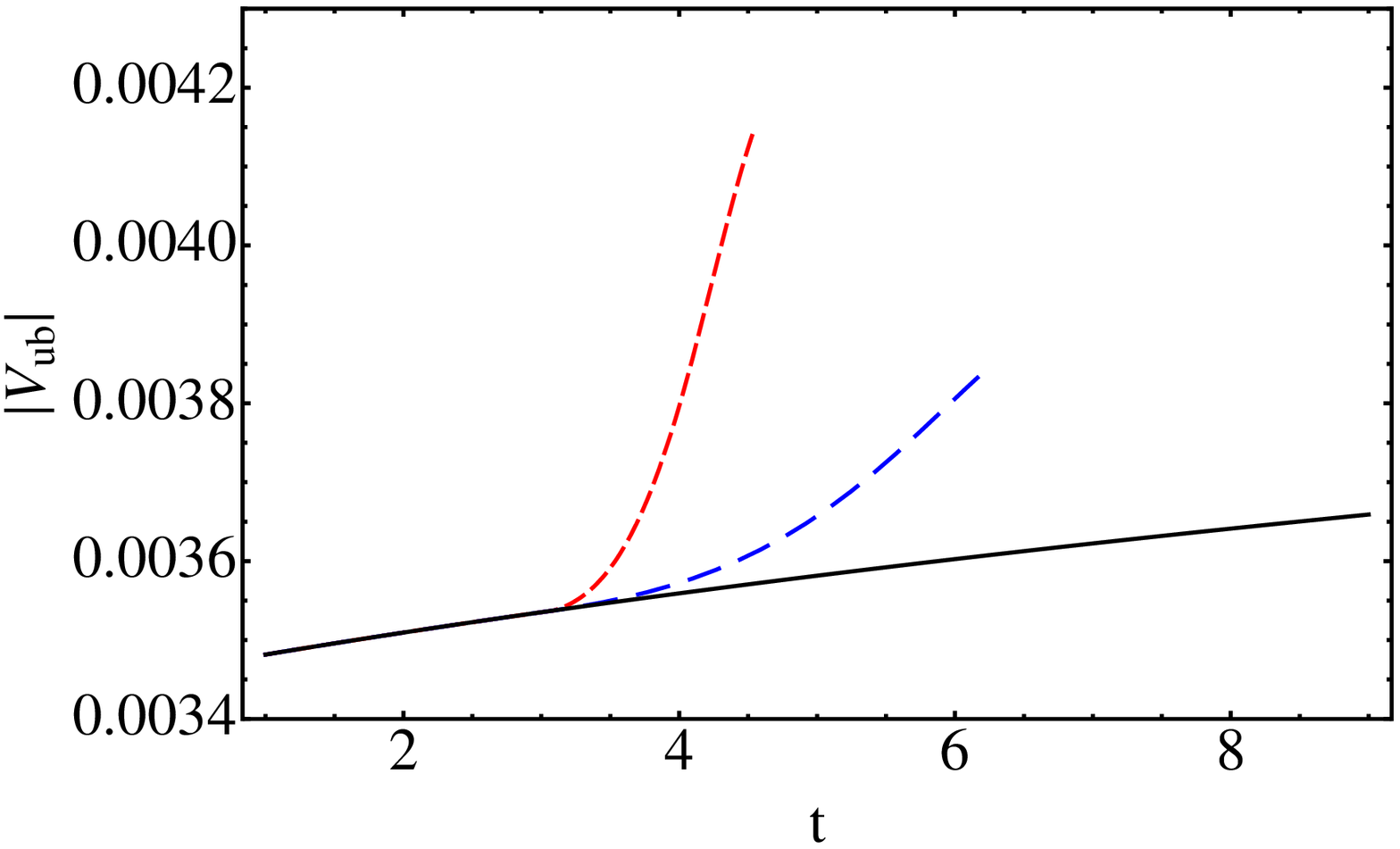} \qquad
\includegraphics[width=7cm,angle=0]{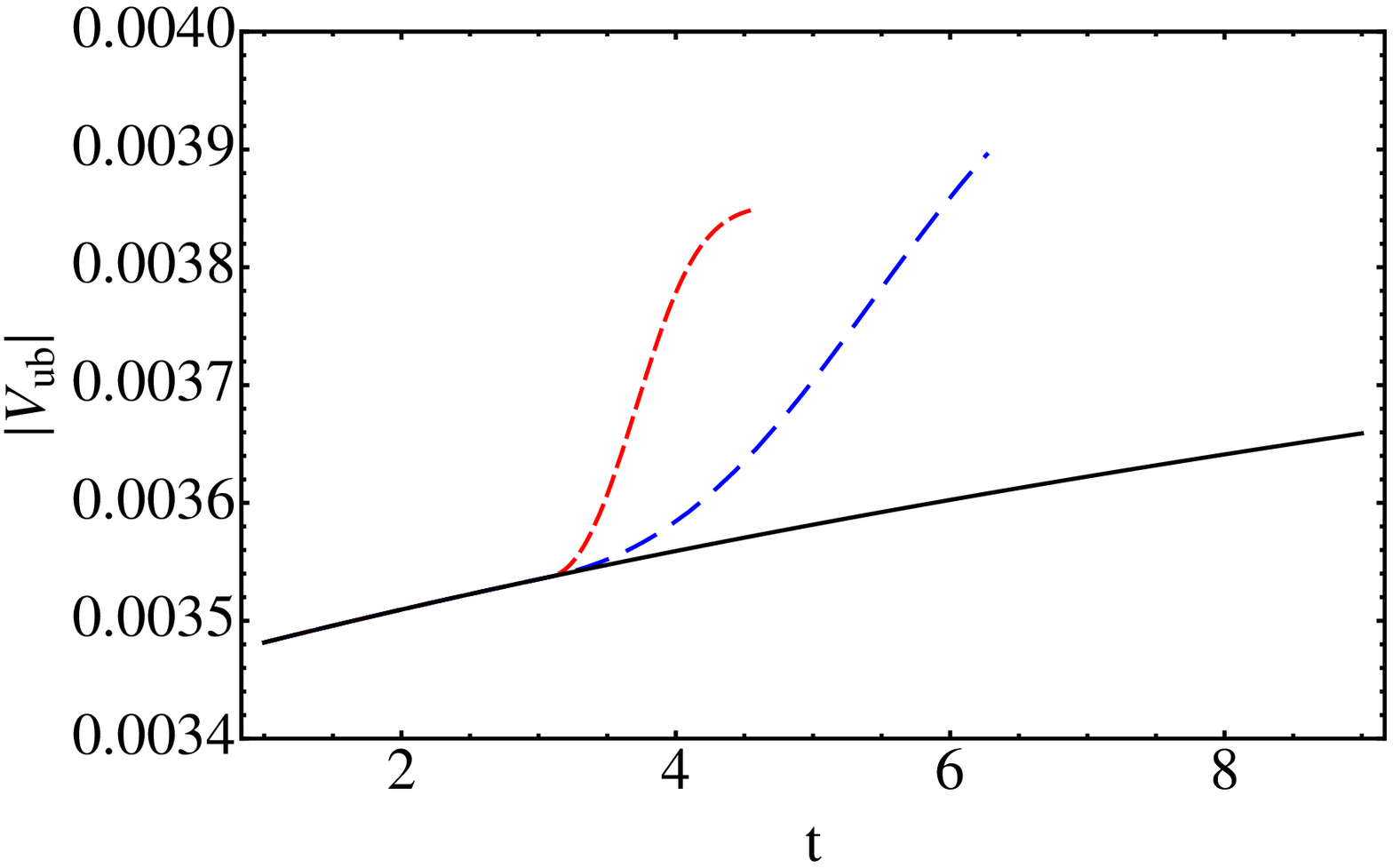}
\end{center}
\caption{ \it (Colour online) Comparison of the $|V_{ub}|$ evolution between the 1UED case (blue) and the 2UED case (red), where the solid line represents the SM case with: in the left panel all matter fields in the bulk; and the right panel for all matter fields on the brane; for a compactification scale of 2 TeV as a function of the scale parameter {$t$}.}
\label{Vub-comparison}
\end{figure}

\par We next turn our attention to the quark flavour mixings, where due to the arbitrary choice of phases for the quark fields, the phases of individual matrix elements of $V_{CKM}$ are not themselves directly observable. We therefore use the absolute values of the matrix element $|V_{ij}|$ as the independent set of rephasing invariant variables. Of the nine elements of the CKM matrix, only four of them are independent, which is consistent with the four independent variables of the standard parameterisation of the CKM matrix.
 
\par We plot in Fig.\ref{Vub-2ued} the evolution of the CKM parameter $|V_{ub}|$ in the bulk and brane cases and note that the other CKM parameters have similar behaviours. We see that once the KK threshold is reached, we have new contributions from the new KK states resulting in a rapidly increasing evolution of the parameter in both cases. Recall that the range of validity for the brane case is bigger than the bulk one, where both cases have a smaller range of validity in the 2UED model when compared to the 1UED model due to the cut-off in two extra dimensions being smaller. For comparison see Fig.\ref{Vub-comparison}.

\begin{figure}[h]
\begin{center}
\includegraphics[width=7cm,angle=0]{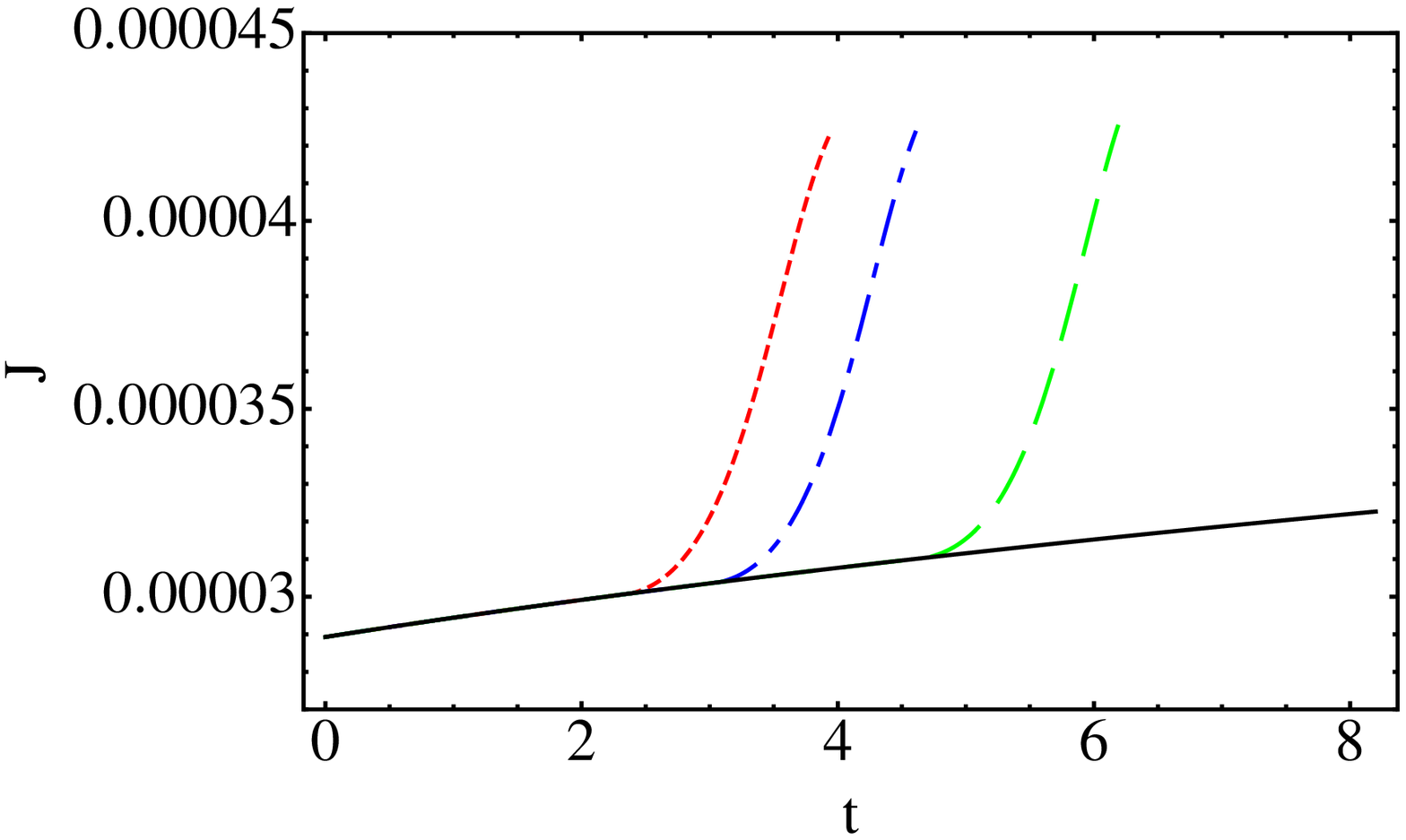} \qquad
\includegraphics[width=7cm,angle=0]{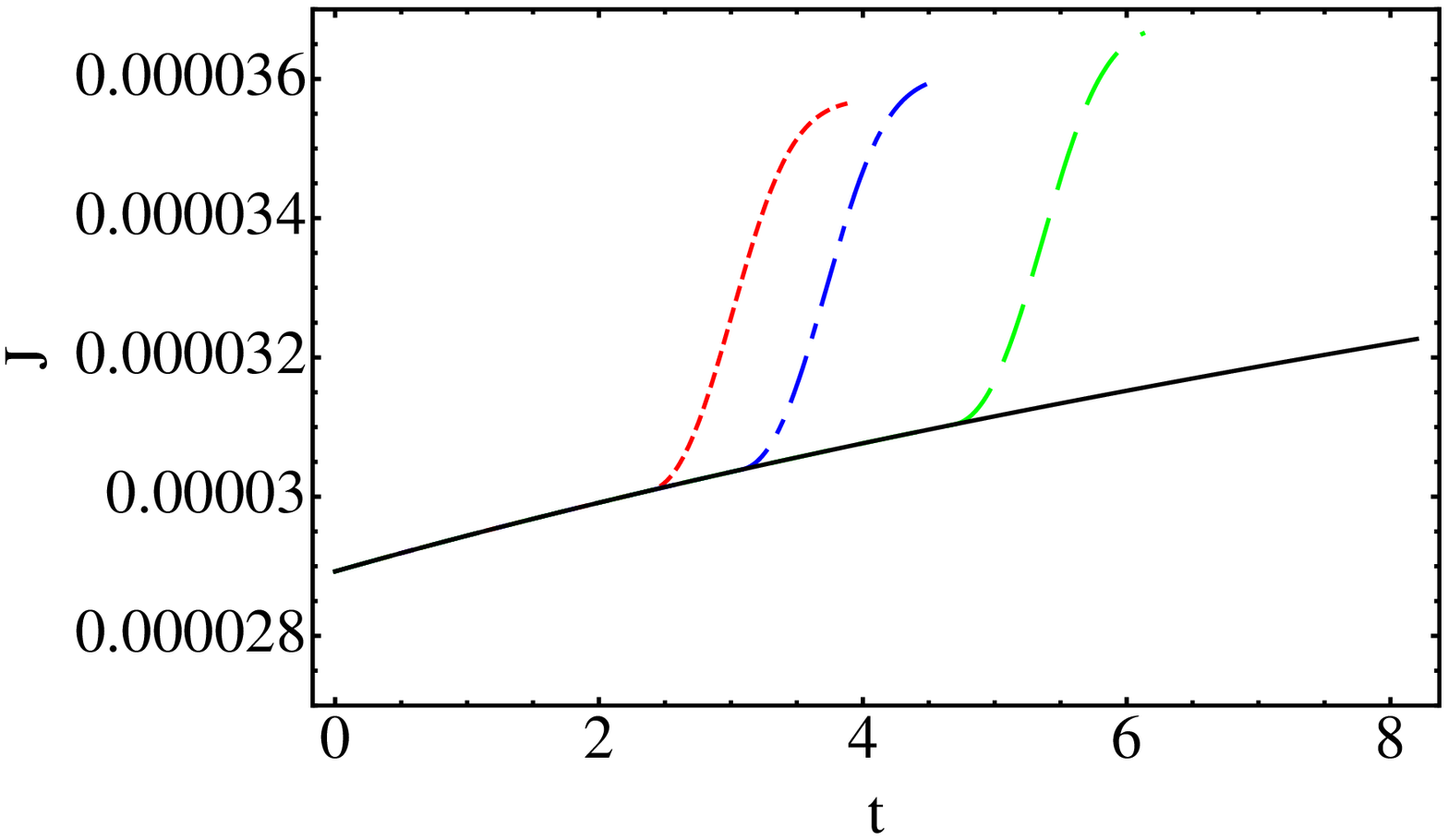}
\end{center}
\caption{ \it (Colour online) The evolution of the Jarlskog parameter $J$ where the solid line represents the SM case with: in the left panel all matter fields in the bulk; and the right panel for all matter fields on the brane; for three different values of the compactification scales 1 TeV (solid line), 2 TeV (dot-dashed line) and 10 TeV (dashed line), as a function of the scale parameter {$t$}.}
\label{J-2ued}
\end{figure}
\begin{figure}[h]
\begin{center}
\includegraphics[width=7cm,angle=0]{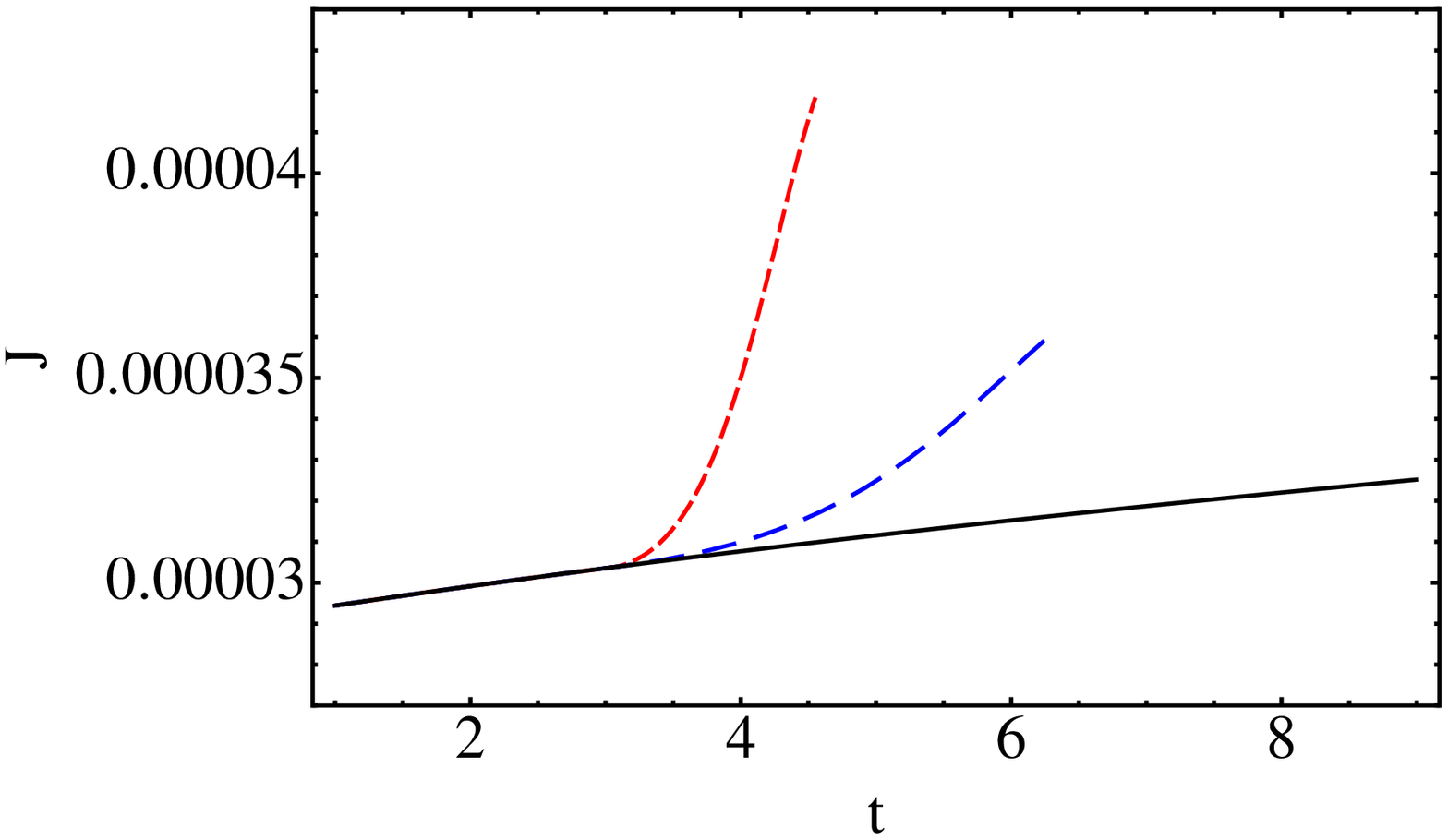} \qquad
\includegraphics[width=7cm,angle=0]{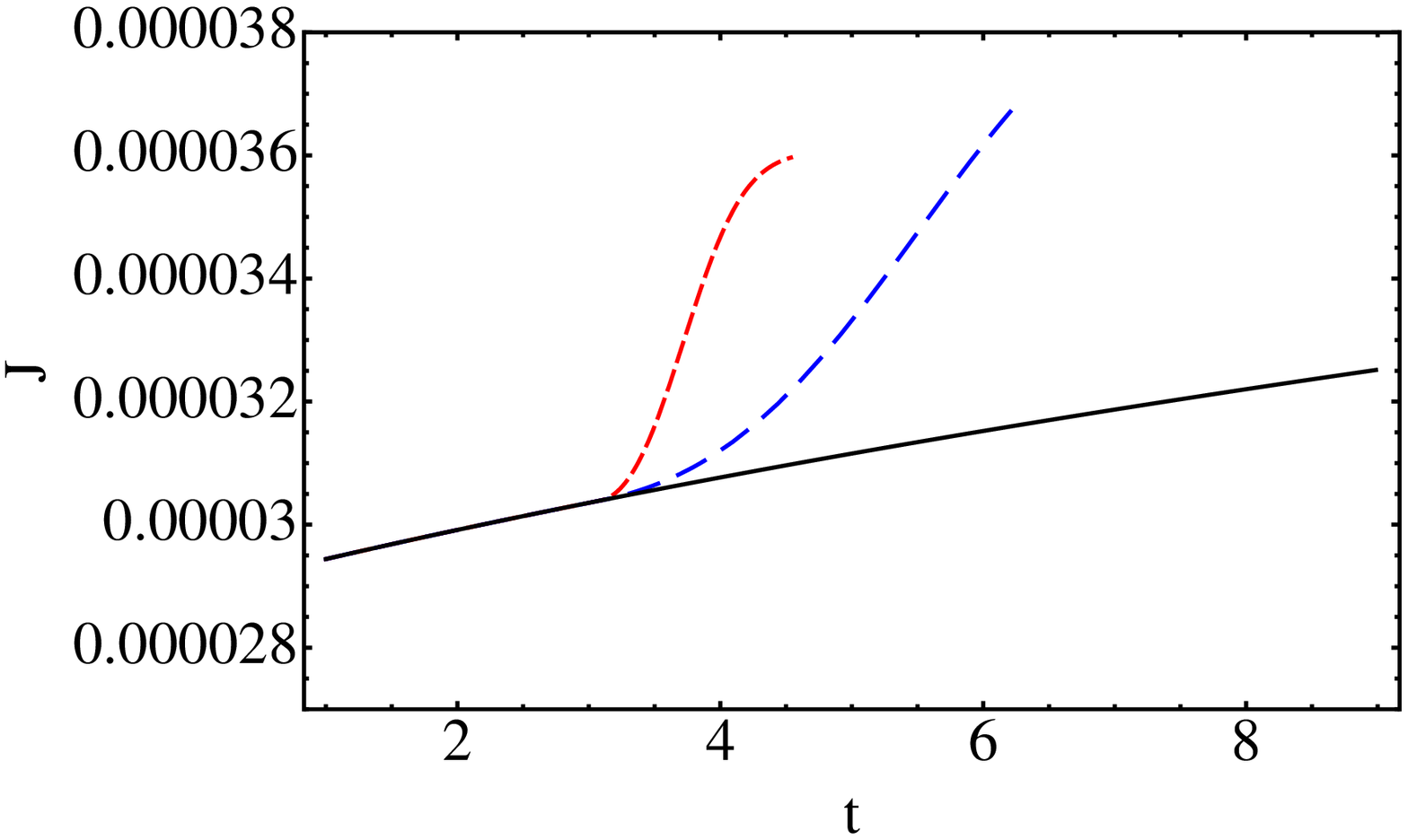}
\end{center}
\caption{ \it (Colour online) Comparison of the Jarlskog parameter $J$ evolution between the 1UED case (blue) and the 2UED case (red), where the solid line represent the SM case with: in the left panel all matter fields in the bulk; and the right panel for all matter fields on the brane; for a compactification scale of 2 TeV as a function of the scale parameter {$t$}.}
\label{J-comparison}
\end{figure}

\par We plot the Jarlskog parameter in Fig.\ref{J-2ued} in the 2UED model for both cases considered here for different radius of compactification. The Jarlskog rephasing invariant parameter $J = \mathrm{Im} V_{ud}V_{cs}V^*_{us} V^*_{cd}$, gives us an indication of the amount of CP violation in the quark sector. As can be seen from the figure, once the first KK threshold is crossed, we have a sharp increase in the value of J up to the cut-off scale for both cases. For the bulk case as approximately 45\%, and the brane localised of 20\%. In Fig.\ref{J-comparison} we compare the 2UED to the 1UED model, and observe similar phenomenologies as for the $|V_{ub}|$ evolution. Note that the main difference between the two models (1UED and 2UED) is the cut-off scale, which for $R^{-1}=1$ TeV is $\Lambda \sim 25$ TeV in the 5D model, which is larger than in the 6D model where $\Lambda \sim 4.5$ TeV. Therefore, the typical 2UED model can be tested, detected or ruled out more easily.
 
\section{Conclusions}
\label{conclusiONS}

\par In this work we derived the RGEs for Yukawa and gauge couplings in the general 2UED model for different scenarios, that of, all matter fields propagating in the bulk or constrained to the brane. We observed that the physical observables in this model undergo rapid evolutions once the first KK threshold in crossed. However, in comparison with 1UED models, we find that this model is valid up to energy scale less than that of the 1UED model cases. This should lead to a means of distinguishing these two models. Note that the case of two extra spatial dimensions opens up a range of different compactification scenarios, as discussed in Appendix \ref{Ssq-factor}, where we have found that in the general (all KK modes included) 2UED models the leading behaviour as encompassed in the $S^2(t)$ dominates. Indeed the fact that only few of the first KK modes are absent has little effect on the numerical results, thus allowing to have robust predictions, reducing considerably the impact of model dependence of the evolution equations and the results described in the present paper.

\section*{Acknowledgements}

AT would like to thank ASC, AA and the NITheP for their hospitality during his stay in Johannesburg, where this work was completed. This work is supported by the National research Foundation (South Africa) and by Campus France (Project Protea-29719RB). We also acknowledge partial support from the Labex-LIO (Lyon Institute of Origins).

\appendix 

\section{The number of KK states in 2UED models} 
\label{Ssq-factor}

\subsubsection*{$T^2$ case}

\par In the $T^2$ case the KK mass is $M_{j,k}= \frac{\sqrt{j^2+k^2}}{R}$, where by define $M_{KK}= \frac{1}{R}$ as the lightest KK mass, the number of KK states originate from:

 \begin{equation}
 \sum_{j,k}{\ln \frac{\mu}{\sqrt{j^2 + k^2} M_{KK}}} \qquad \mbox{for}  \qquad 1\leq j^2 + k^2 \leq \left(\frac{\mu}{M_{KK}}\right)^2
 \end{equation}
 \begin{equation}
 \sum_{j,k}{\ln \frac{\mu}{\sqrt{j^2 + k^2} M_{KK}}} = \sum_{j,k}{\ln \frac{\mu}{M_{KK}}-\frac{1}{2} \ln(j^2 + k^2 )}\;,
 \end{equation} 
\noindent One can use the polar coordinates and change the sum into an integral as 
\begin{equation}
\sum_{j,k}{\ln \frac{\mu}{M_{KK}}-\frac{1}{2} \ln(j^2 + k^2)} = \int^{\frac{\mu}{M_{KK}}}_{1} { 2 \pi r dr  \left(\ln \frac{\mu}{M_{KK}}- \ln r \right)}\;,
 \end{equation}
from which we obtain 
  \begin{equation}
 \frac{\pi}{2} \left[\left(\frac{\mu}{M_{KK}}\right)^2 -1 - 2 \ln \frac{\mu}{M_{KK}}\right]\;.
  \end{equation}
 
\par Therefore, the gauge couplings running equation becomes
 \begin{equation}
 \frac{4 \pi}{g^2_i(\mu)}= \frac{4 \pi}{g^2_i(M_Z)}- \frac{b^{SM}_i}{2 \pi} \ln\frac{\mu}{M_Z} + 2 C \frac{b^{6D}_i}{2 \pi} \ln\frac{\mu}{M_{KK}}- C\frac{b^{6D}_i}{2 \pi} ((\frac{\mu}{M_{KK}})^2 -1)\;, \label{app-5}
 \end{equation}
 where $C= \frac{\pi}{2}$, and in terms of the $t$ parameter we have $\mu= M_Z e^t$. As such Eq.(\ref{app-5}) becomes
 \begin{equation}
 \frac{4 \pi}{g^2_i(\mu)}= \frac{4 \pi}{g^2_i(M_Z)}- \frac{b^{SM}_i}{2 \pi} \ln\frac{M_Z e^t}{M_Z} + 2 C \frac{b^{6D}_i}{2 \pi} \ln\frac{M_Z e^t}{M_{KK}}- C\frac{b^{6D}_i}{2 \pi} \left(\left(\frac{M_Z e^t}{M_{KK}}\right)^2 -1\right)\;. \label{app-6}
 \end{equation}
 
\par We take the derivative of Eq.(\ref{app-6}) with respect to $t$ to obtain
 \begin{eqnarray}
 - 2g^{-3}_i \frac{d g_i}{dt} &=& - \frac{b^{SM}_i}{8 \pi^2} + 2 C \frac{b^{6D}_i}{8 \pi^2}\left( 1- \left(M_Z R e^t\right)^2\right) \nonumber \;, \\
  \frac{d g_i}{dt} &=&  \left[\frac{b^{SM}_i}{16 \pi^2} + 2 C \frac{b^{6D}_i}{16 \pi^2} \left(S(t)^2 -1\right)\right]g^3_i \;, \label{app-7}
 \end{eqnarray}
 \noindent which is Eq.(\ref{gauge2UED}). Thus our KK number for the general 2UED model is $2C( S(t)^2 -1)$, where our general model is this $T^2$ model, and where $S(t)= M_Z R e^t$ assuming that all modes contribute in the range of our energy scale.
 
 \subsubsection*{$S^2$ case}
 
\par In the $S^2$ case the KK mass is $M_{j,k}= \frac{\sqrt{j(j+1)}}{R}$, and we have $M_{KK}= \frac{\sqrt{2}}{R}$ being the lightest KK state in this model. Our KK number is then given by 
\begin{equation}
 \sum_{j,k} (2j+1){\ln \frac{\mu}{\sqrt{j^2 + k^2} M_{KK}}} \qquad  \mbox{for}\;\;  2\leq j(j+1) \leq 2(\frac{\mu}{M_{KK}})^2 \;,
 \end{equation}
where $2j+1$ is the number of degenerate states in each j-level. 
 
\par Using $j(j+1)\approx j^2$ and defining $j_{max} = \sqrt{2}\frac{\mu}{M_{KK}}$ and  $j_{min}= \sqrt{2}$, 
\begin{equation}
\sum_{j,k}{\ln \frac{\mu}{\sqrt{\frac{j(j+1)}{2}} M_{KK}}} = \int^{\sqrt{2}(\frac{\mu}{M_{KK}})}_{\sqrt{2}} { 2 j  dj\ln \frac{j_{max}}{j}}\;,
\end{equation}
and
\begin{equation}
\int^{\sqrt{2}(\frac{\mu}{M_{KK}})}_{\sqrt{2}} 2 j  dj\ln \frac{j_{max}}{j} \approx \left[\left(\frac{\mu}{M_{KK}}\right)^2 -1 - 2 \ln \frac{\mu}{M_{KK}}\right]\;.
\end{equation}
Therefore our KK number, as function of the $t$ parameter, in this model is given by $2(S(t)^2 -1)$ assuming  that all modes contribute in the range of our energy scale.
 
\subsubsection*{Model dependence of the RGE}

\par Note that, in specific realisations of the 2UED models some of the states may not be present, therefore one needs to subtract those states which do not contribute from the total KK number. For instance the case of $T^2$ or $S^2$ compactifications, the states (0,2k) and (2k,0) for a given parity may not be present. Assuming that these states are not there, the number of KK states for a such model becomes $2C(S^2-1)-2(S-1)$, where $C=\frac{\pi}{2}$ for the $T^2$ case or 1 for the $S^2$ case. We have numerically analysed such 2UED models to test the model dependence of the results obtained with the RGE and only minor changes in the plots were observed, with the major phenomenology discussed in section \ref{resulTS} remaining unaltered. 


\end{document}